\documentclass[11pt, letter]{article}
\usepackage{jheppub}
\usepackage[english]{babel}
\usepackage{blindtext}\blindmathtrue
\usepackage{avant} % Use the Avantgarde font for headings
\usepackage[dvipsnames]{xcolor}
\usepackage{amsmath,epsf,amssymb,mathtools,latexsym,amsthm,setspace,array,pifont,hyperref,amsfonts,dsfont,cancel,braket,parskip}
\usepackage[none]{hyphenat} % No hyphenated line endings
\usepackage{graphicx}
\usepackage{float}
\usepackage{tikz}
\usepackage{tikz-feynman, contour}
\tikzfeynmanset{compat=1.1.0}
\usetikzlibrary{shapes,arrows,positioning,automata,backgrounds,calc,er,patterns}
\makeatletter
\newlength\CoolS@sizex
\newlength\CoolS@sizey
\newcommand*\CoolS@inner{%
\begin{tikzpicture}[baseline=0.04\CoolS@sizey]%
\foreach \x in {0, 1, ..., 5} \foreach \y in {0, 1, ..., 10}
\coordinate (c\x\y) at (\x *0.12*\CoolS@sizex, \y *0.107*\CoolS@sizey);
\draw [line width=\Cool@stroke] (c28)--(c26)--(c44)--(c42)--(c20)--(c02)--(c04)--(c15);
\draw [line width=\Cool@stroke] (c22)--(c24)--(c06)--(c08)--(c210)--(c48)--(c46)--(c35);
\end{tikzpicture}}
\usepackage{todonotes}
\graphicspath{{Pictures/}} % Specifies the directory where pictures are stored
\newcommand{\nn}{\nonumber}
\newcommand{\sd}{\mathrm{d}}
\newcommand{\pd}{\partial}
\newcommand{\da}{\dot{a}}
\newcommand{\db}{\dot{b}}

\newcommand{\dA}{\dot{\alpha}}
\newcommand{\dB}{\dot{\beta}}

\newcommand{\A}{\alpha}
\newcommand{\B}{\beta}
\newcommand{\R}{\mathbb{R}}

\newcommand{\cl}[1]{\mathcal{#1}}

\def\prd{\ref@{Phys.~Rev.~D}}        % Physical Review D
\newcommand{\Tr}[1]{\text{Tr}\left(#1\right)}
\usepackage{tcolorbox}
\definecolor{airforceblue}{rgb}{0.36, 0.54, 0.66}
\definecolor{azure}{rgb}{0.0, 0.5, 1.0}
\newtcolorbox{tdbox}{colback=airforceblue!40!white,colframe=azure!90!black} 
\newcommand{\td}[1]{
	\if\notesOn1
	\begin{tdbox}
		#1
	\end{tdbox}
	\fi
}
\hypersetup{
	colorlinks = true,
	linkcolor = Mahogany,
	anchorcolor = Mahogany,
	citecolor = Mahogany,
	filecolor = Mahogany,
	urlcolor = Mahogany
}

\def\notesOn{1}
\tikzset{
	graviton/.style={
		double,
		decoration={snake, aspect=0.75, mirror, segment length=1.5mm},
		decorate
	}
}

\tikzfeynmanset{
	bigblob/.style={
		shape=circle,
		draw=blue,
		fill=red}
}

\usepackage{enumitem,amssymb}
\newlist{todolist}{itemize}{2}
\setlist[todolist]{label=$\square$}
\usepackage{pifont}

\renewcommand{\textcolor}[2]{#2}
\title{Anyons and the Double Copy}
\author[1] {Daniel J Burger,}
\author[2]{William~T.~Emond}
\author[3]{and~Nathan Moynihan}
\affiliation[1]{The Laboratory for Quantum Gravity \& Strings, Department of Mathematics \& Applied Mathematics, University Of Cape Town}
\affiliation[2]{CEICO, Institute of Physics of the Czech Academy of Sciences, Na Slovance 2, 182 21 Praha 8, Czech Republic}
\affiliation[3]{Higgs Centre for Theoretical Physics, School of Physics and Astronomy, The University of Edinburgh, EH9 3FD, Scotland}
\emailAdd{burgerjdaan@gmail.com}
\emailAdd{william.emond@fzu.cz}
\emailAdd{nathantmoynihan@gmail.com}
\abstract{
We examine the double copy structure of anyons in gauge theory and gravity. Using on-shell amplitude techniques, we construct little group covariant spinor-helicity variables describing massive particles with spin, which together with locality and unitarity enables us to derive the long-range tree-level scattering amplitudes involving anyons. We discover that classical gauge theory anyon solutions double copy to their gravitational counterparts in a non-trivial manner. Interestingly, we show that the massless double copy captures the topological structure of curved spacetime in three dimensions by introducing a non-trivial mixing of the topological graviton and the dilaton. Finally, we show that the celebrated Aharonov-Bohm phase can be derived directly from the constructed on-shell amplitude, and that it too enjoys a simple double copy to its gravitational counterpart.
}

\begin{document}
\maketitle
\section{Introduction}
Particles with fractional statistics --- anyons --- are a peculiarity unique to planar physics, where we restrict our attention to an effective 2+1 dimensional world. Anyons are pervasive in condensed matter physics \cite{Wilczek:1981du}, naturally explaining the (experimentally observed) fractional quantum hall effect \cite{Halperin:1984fn}, high temperature superconductivity \cite{Laughlin:1988fs} and the physics of cosmic strings \cite{Cho:1992wu}. Recently, they are of renewed interest due to them being the central component in topological quantum computing, a leading candidate for building a functioning quantum computer. Anyons carry both electric charge and magnetic \textit{flux}, and can be thought of as the 2+1 dimensional cousin of the \textit{dyon}, which carry both electric and magnetic charges and have been recently studied using on-shell amplitudes \cite{Emond:2020lwi,Moynihan:2020gxj,Huang:2019cja,Kim:2020cvf,Terning:2020dzg,Csaki:2020inw}. Over the last two decades, on-shell scattering amplitude techniques have been highly successful in enhancing our understanding of gravitational physics, becoming standard tools to understand the two-body problem \cite{Bern:2019nnu,Bern:2019crd,Kalin:2020fhe,Kalin:2020lmz,Kalin:2020mvi,Kalin:2019rwq,Kalin:2019inp,Mogull:2020sak,Bern:2021dqo,Bern:2020uwk,Jakobsen:2021smu}, utilising many of the simplifications that on-shell philosophy has to offer, such as the double copy \cite{Bern:2008qj} (including in massive gravity \cite{Johnson:2020pny,Momeni:2020vvr,Momeni:2020hmc}). In this paper, we would like to apply on-shell techniques to systems usually associated with condensed matter physics, although, as we will see, this will also have a direct implications for gravitational systems via the double copy.

Anyons are induced by minimally coupling point particles to a `statistical' gauge boson $A_\mu$, whose dynamics is governed by the Chern-Simons action. Among other things, this means that anyons are solutions to topologically massive gauge theories, which were recently conjectured by one of us to double copy to topologically massive gravity (TMG) \cite{Moynihan:2020ejh}. If this is true, then anyon solutions ought to double copy as well, and in this paper we will investigate whether or not this is the case.

\begin{figure}[h!]\centering\label{fig:dcgraph}
	\begin{tikzpicture}[every node/.style={midway}]
		\matrix[column sep={24em,between origins}, row sep={12em}] at (0,0) {
			\node(E) {\small{Massive vector}}  ; & \node(M) {\small{Massive graviton $\oplus$ Massless spin-2 ghost}}; \\
			\node(3DE) {\small{Massless vector}}; & \node (3DCS) {\small{Dilaton $\oplus$ Massless graviton}};\\
		};
		\draw[->] (E) -- (M) node[anchor=south]  {Double Copy};
		\draw[->] (E) -- (3DE) node[anchor=east,xshift=-0.4cm,yshift=-0.8cm,rotate=-90]  {$m \longrightarrow 0$};
		\draw[->] (3DE) -- (3DCS) node[anchor=south] {Double Copy};
		\draw[->] (M) -- (3DCS) node[anchor=west,xshift=0.4cm,yshift=0.8cm,rotate=-90] {$m \longrightarrow 0$};
	\end{tikzpicture}\caption{The massive and massless 2+1 D double copy. All massive particles have purely topological mass.}
\end{figure}
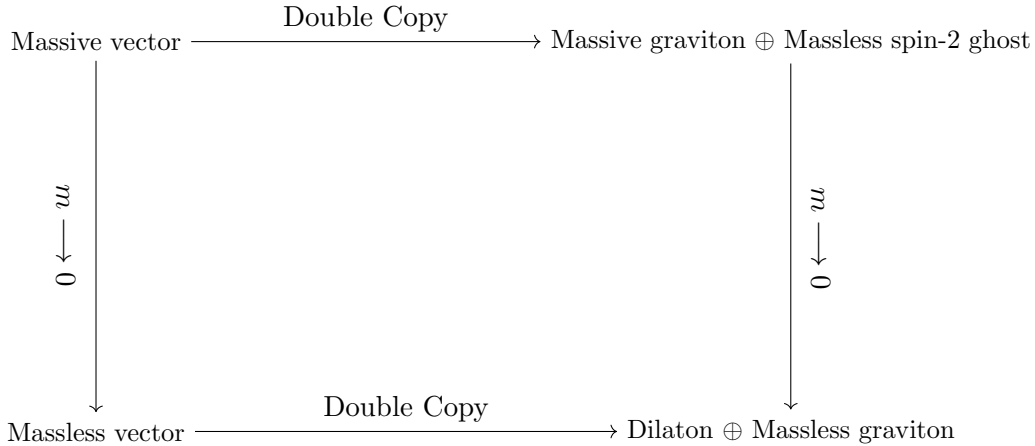

We find that the double copy does indeed hold, but in a fairly non-trivial way: at tree-level, the double copy of a topologically massive gauge boson leads to a combination of the entirely topological solution from Einstein gravity and the topologically massive graviton. This is perhaps not too surprising, since the spectrum of TMG contains a massless ghost graviton in addition to a massive graviton, but what is surprising is that the double copy contains an \textit{extra} ghost graviton. The full structure of the massive and massless double copy is summarised in Fig. \ref{fig:dcgraph}. We note that massless gravitons in 2+1 dimensions are entirely topological, so we are not gaining any extra degrees of freedom by adding additional massless gravitons via the double copy.

What is perhaps more surprising is that in the massless limit we find that the extra topological graviton is crucial to ensuring that the classical double copy is obtained, as found in Ref. \cite{CarrilloGonzalez:2019gof} using a Kerr-Schild construction. We find that the classical double copy solution discovered there is a mixture of this topological graviton and the dilaton.

 At tree-level, then, the massless 3D double copy appears to hold in the standard way, in the sense that we can schematically write
\begin{equation}
	A_\mu\otimes A_\nu \sim h_{\mu\nu}\oplus\phi \;,
\end{equation}
which holds even when there are no propagating gravitons. We note that on-shell, massive gravitons and massive/massless vectors both have one degree of freedom, whereas the massless graviton has none.

\section{Anyons in Gauge Theory and Gravity}\label{sec:anyons}
In this section, we will establish some basic properties of both relativistic and non-relativistic anyons in quantum field theory and gravity. In the case of a quantum field theory, we will also include a standard Maxwell kinetic term giving an action
\begin{equation}\label{EMaction}
	S =  -\int \sd^3x~\left(\frac{1}{4}F_{\mu\nu}F^{\mu\nu} - \frac{ke^2}{2}\epsilon^{\mu\nu\rho}A_\mu F_{\nu\rho}\right) + S_{matter}[\phi;A_\mu] \;,
\end{equation}
where $[A_\mu]=1/2$, $[e] = 1/2$, $[k] = 0$ and $k$ is the Chern-Simons level number that determines the anyonic statistics. Note that we work in the $(-++)$ signature throughout this discussion. 

We briefly remark here on the topological nature of such a theory. Indeed, it can shown that the Chern-Simons (CS) contribution to this action is entirely independent of the choice of metric, relying solely on the underlying properties of the spacetime manifold. This sector of the theory thus furnishes a description based purely on the topology of the manifold it is embedded in. Moreover, by combining it with the standard massless sector of a given gauge theory (in this case electromagnetism) it is possible to generate a gauge invariant mass term, rendering the gauge field ``topologically'' massive, albeit manifestly breaking parity and time-reversal symmetry (for further details we refer the reader to Ref.~\cite{Deser:1981wh}).
	
This analysis can be readily applied to gravity as well, in which one can include an analogous gravitational CS action. All of the properties observed above carry over, although an interesting observation in this case is that pure general relativity is trivial in 2+1 dimensions (i.e., there are no local degrees of freedom). However, the addition of a CS piece dramatically changes the picture, introducing a topologically massive graviton, and with it, non-trivial solutions. We shall explore this in detail later on.

Returning to the discussion at hand, let us consider the action~\eqref{EMaction}, where matter is minimally coupled to the gauge field $A_\mu$ by a massive, charged matter current $j^\mu = (\rho,\mathbf{j})$, where $\rho$ and $\mathbf{j}$ are the charge density and current density respectively, and $[j^\mu]=5/2$ (the fractional mass dimension here arises due to the elementary charge $e$ having mass dimension $[e]=1/2$, as opposed to the 4D case where it is dimensionless)\footnote{Here, our discussion draws upon details in chapter 3 of Ref.~\cite{lerda2008anyons}.}. The corresponding field equations are then given by 
\begin{equation}\label{eq:MCS eom}
	\pd_\nu F^{\mu\nu} + \frac{ke^2}{2}\epsilon^{\mu\nu\rho}F_{\nu\rho} = j^\mu \;.
\end{equation}
Moreover, by varying the action~\eqref{EMaction} under a $U(1)$ gauge transformation $A_\mu\rightarrow A_\mu + \partial_\mu\Lambda$ (where $\Lambda$ is a spacetime dependent gauge parameter), it can be shown that the associated Noether current is given by  
\begin{equation}\label{eq:MCS conserved current}
	J^\mu = j^\mu - \frac{ke^2}{2}\epsilon^{\mu\nu\rho}F_{\nu\rho} \;.
\end{equation}
Note that both terms in $J^\mu$ are actually individually conserved by virtue of the continuity equation for $j^\mu$ and the Bianchi identity for the Chern-Simons current. Interestingly, the Chern-Simons current arises precisely because the corresponding action varies by a total derivative under a gauge transformation. Clearly, such a term is not present in the conserved current found in pure Maxwell theory, due to its manifest gauge invariance. 

Following on from this, by integrating the zeroth component of eq.~\eqref{eq:MCS conserved current} over a small 2D disk containing a single anyon, we can determine the conserved charge associated to it:
%\footnote{Note that Q has mass dimension $[Q]=1/2$ and so is proportional to $e$, however, the charge density $\rho$ has mass dimension $[\rho]=2$ and so $\int\sd^2x\,\rho = e^2$ as opposed to $e$ as it would be naively expected in analogy the 4D result.}:
\begin{equation}
	Q = \int\sd^2x\,J^0 = \int\sd^2x\,\big(\rho - ke^2\epsilon^{ij}\partial_i A_j\big) = e\big(1 - k\Phi\big) \;,
\end{equation}where $\Phi$ is the magnetic flux through the 2D disk enclosing the anyon, defined as
\begin{equation}
	\Phi = \int \sd^2x~\epsilon^{ij}\pd_i A_j \;.
\end{equation} 
We see then that anyons carry both electric charge $e$ and magnetic flux $\Phi$, with the product of them induced by the Chern-Simons term in the action. This is often called `flux attachment' in the literature, since the introduction of a Chern-Simons term has meant that we have attached a magnetic flux to each charged particle. Adding flux to a charged particle has an interesting effect on its angular momentum: it shifts it by an amount proportional to the flux \cite{Wilczek:1981du}. In particular, for a pure Chern-Simons interaction, i.e., eq.~\eqref{eq:MCS eom} without the Maxwell contribution $\partial_\nu F^{\mu\nu}$, it follows that $\epsilon^{ij}\partial_iA_j = \rho/e^2k$ and hence $\Phi = 1/ek$.

If we consider an infinitely thin solenoid containing some flux along $z$, perpendicular to our 2+1 dimensional plane, then the potential outside the solenoid does not vanish. The magnetic field is restricted to be within the solenoid, and is proportional to a delta function at the location that it intersects the perpendicular plane (which we choose to be the origin), given by
\begin{equation}\label{magfield}
	B(r) = \frac12\frac{\delta\left(\epsilon^{\mu\nu\rho}A_{\mu}F_{\nu\rho}\right)}{\delta A_0} = \epsilon^{ij}\pd_i A_j \equiv \Phi\delta(r) \;.
\end{equation}
Choosing an ansatz of the form $A_i = \epsilon_{ij}\pd^j f(r)$ gives a differential equation which can be solved to find a gauge field written in terms of the flux as
\begin{equation}
	A_i = \frac{\Phi}{2\pi}\epsilon_{ij}\pd^j\log(r) \;.
\end{equation}

The angular momentum is now defined as $\textbf{L} = \textbf{r}\times \left[\textbf{p} -e\textbf{A}\right]$. Working in polar coordinates, rotations in the $x-y$ plane by an angle $\varphi$ around the solenoid will now be generated by the angular momentum operator
\begin{equation}\label{eq:am operator}
	\ell_z = -i\pd_\varphi -eA_\varphi \;,
\end{equation}
where the additional contribution comes from the fact that the gauge potential is non-zero outside $r=0$. In polar coordinates, we find that only the $\varphi$ component of the gauge field is non-zero, and given by
\begin{equation}
	A_\varphi = \frac{\Phi}{2\pi} \;.
\end{equation}
Given this, we can then consider the effects of such an external electromagnetic field on the quantum mechanics of a charged particle (of mass $m$). In the time-independent case, in which the particle is subject only to a (time-independent) electromagnetic potential $A^\mu(\mathbf{r})=\left(\phi(\mathbf{r}),\mathbf{A}(\mathbf{r})\right)$, the corresponding Schr\"{o}dinger equation is given by $\frac{1}{2m}\left(i\hbar\nabla^2+\frac{q}{c}\mathbf{A}(\mathbf{r})\right)^2\psi(\mathbf{r})+q\phi(\mathbf{r})\psi(\mathbf{r}) = E\psi(\mathbf{r})$. Noting that the wavefunction $\psi(\mathbf{r})$ describing the charged particle is separable, and thus by working in a polar coordinate system $(r,\varphi)$, it can be shown that the angular part (dependent on $\varphi$) is given by
\begin{equation}
	\psi_n(\varphi) \propto e^{in\varphi} \;.
\end{equation}
Accordingly under a rotation (generated by \eqref{eq:am operator}) we find that the wavefunction becomes
\begin{equation}\label{anyonspin}
	\ell_z\psi(r,\varphi) = \left(n - \frac{e\Phi}{2\pi}\right)\psi(r,\varphi) \;,
\end{equation}
i.e. the eigenvalue of the angular momentum operator is shifted by an amount proportional to the flux, and is no longer required to be an integer. Under a gauge transformation $A_i' = A_i + \pd_i\Lambda$, the wavefunction picks up a phase of the form
\begin{equation}\label{newphase}
	\psi'(x) = e^{ie\Lambda(x)}\psi(x) \;.
\end{equation}

We can actually gauge away the contribution from $A_\varphi$ in the angular momentum operator by doing a singular gauge transformation of the form
\begin{equation}
	A_i \rightarrow A'_i = A_i + \pd_i\Lambda \;,
\end{equation}
provided we choose the gauge function such that it satisfies $\pd_i\Lambda = -A_i$. We then find that $A'_i = 0$ and covariant derivative is simply the flat space derivative and all particles are free; this is the so-called `anyon gauge'. Intriguingly, despite this gauge choice describing free particles, the transformation required is topologically non--trivial, resulting in the wavefunction picking up an observable (quantum) phase generated by the gauge transformation according to eq. \eqref{newphase}, meaning that this phase is gauge invariant\footnote{In fact, it can be shown that the phase depends purely on the geometry of the closed path traversed by a charged particle in the presence of a magnetic field.}. Thus, although we have naively removed the effects of the gauge potential, its presence is still felt through the imposed boundary conditions, i.e. the additional contribution to the angular momentum operator is a physical observable, not simply a gauge artefact. 

To understand the particle statistics in such a set-up we need to extend this analysis to the multi-particle case. It is sufficient to consider a system of two identical (mutually) non-interacting particles, in which we now take $\psi(r,\varphi)$ to denote the relevant part of the two-body wavefunction, in relative coordinates $(r,\varphi)$\footnote{To elaborate a bit, the two-body Hamiltonian in this set-up can be expressed in the centre-of-mass (CM) frame in relative coordinates. The resulting wavefunction is then a product of the CM wavefunction, describing the system as a whole, and the ``relative" wavefunction $\psi(r,\varphi)$, describing the behaviour of the two particles in relation to one another. Here, it is only $\psi(r,\varphi)$ that is relevant for the discussion, since it is sensitive to the particle statistics, whereas the CM wavefunction is not, hence we drop its contribution.}. We can then essentially follow an analogous argument as the one-particle case above. Indeed, it follows that under an exchange of the two particles (an anticlockwise half-rotation\footnote{Note that the wavefunction of two anyons undergoing a full rotation does not give the same wavefunction back, but one multiplied by $e^{i2\theta}$. See e.g. \cite{Wilczek:1981du,Sen:1993qc} for further details.}), the wavefunction becomes
\begin{equation}
	\psi(r,\varphi + \pi) = e^{i\pi\ell_z}\psi(r,\varphi) = e^{i\left(\pi n - \frac{1}{2}e\Phi\right)}\psi(r,\varphi) \;.
\end{equation}
When the flux vanishes, we see that the wavefunction obeys the usual bosonic ($n$ an integer) or fermionic ($n$ a half-integer) statistics. However, when $\Phi \neq 0$ the phase can be entirely arbitrary: the wavefunction obeys anyonic statistics. For a Chern-Simons interaction, the flux is given by $\Phi = \frac{\hbar}{ek}$, such that the wavefunction picks up a phase inversely proportional to the level number
\begin{equation}
	\psi(r,\varphi + \pi) = e^{\pi i\ell_z}\psi(r,\varphi) = e^{i\left(\pi n - \frac{1}{2k}\right)}\psi(r,\varphi) \;.
\end{equation}
 
For gravity, there is an analagous effect arising in both Einstein gravity with spinning sources and in topologically massive theories of gravity. For generality, we will consider our system to be described by the action 
\begin{equation}
	S_{TMG} = -\frac{1}{2\kappa^2}\int \sd^3x\sqrt{-g}\left[R + \frac{1}{2m}\epsilon^{\lambda\mu\nu}\Gamma^\rho_{\lambda\sigma}\left(\pd_\mu\Gamma^\sigma_{\nu\rho} + \frac{2}{3}\Gamma^\sigma_{\mu\tau}\Gamma^\tau_{\nu\rho}\right)\right] + S_{matter}[\phi;g_{\mu\nu}] \;,
\end{equation}
where $m$ is the topological mass parameter, and $\kappa^2=8\pi G$. The solutions of this theory are \textit{gravitational anyons} \cite{Deser:1989ri,Ortiz:1991gx}, and they share many of the same properties as their electromagnetic counterparts. Asymptotically, the metric for a gravitational anyon with mass $M$ is given by
\begin{equation}
	\sd s^2 = -(\sd t + 4G\Sigma \sd\theta)^2 + r^{-8GM}(\sd r^2 + r^2\sd\theta) \;,
\end{equation}
where $\Sigma = \sigma + \frac{M}{m}$ is the `spin', with $\sigma$ the spin of the source and the $M/m$ factor is the spin induced by the Chern-Simons term \cite{Deser:1989ri,Ortiz:1991gx,Deser:1990ve}. We see that gravitational anyons necessarily spin, either because the source has intrinsic spin $\sigma$, or because of presence of the Chern--Simons interaction.

If we now consider a neutral massive particle orbiting a spinning gravitational point source, we find that the angular momentum operator has a similar form to the electro-- magnetic case, where now charge-flux has been replaced with energy-spin \cite{deSousaGerbert:1988qzd}
\begin{equation}
	\ell_z = -i\pd_\varphi -E\Sigma \;,
\end{equation}
where $E$ is the energy. We note that there is also interesting physics at the critical value of $M + m\sigma = 0$, where the system reduces to that of a conical singularity \cite{Clement:1990mp}. By performing a singular coordinate transformation of the metric (singular at the origin)
\begin{equation}
	t \rightarrow T - 4G\Sigma\theta \;,~~~~~r\rightarrow \left((1-4GM)R\right)^{\frac{1}{1-4GM}}\;,~~~~~\theta \rightarrow \frac{1}{1-4GM}\Theta \;,
\end{equation}
we find that the metric becomes that of flat space
\begin{equation}
	\sd s^2 = -\sd T^2 + \sd R^2 + R^2\sd\Theta^2 \;.
\end{equation}
While this does describe flat space, we find that again the gauge transformation has meant that our coordinates now have new boundary conditions, inherited from the fact that $\theta$ is identified with $\theta + 2\pi$, such that now we have
\begin{equation}
	(T,\Theta) \mapsto \left(T -8\pi G\Sigma, \Theta + 2\pi(1-4GM)\right) \;.
\end{equation}
We note that gravitational anyons with these boundary conditions exist in Einstein gravity in 2+1 dimensions and do not require the Chern-Simons term (or a massive graviton), so long as there is a spinning source. To see this, we simply note that the infinite mass limit, which gives back Einstein gravity, yields the same boundary conditions with $\Sigma\rightarrow\sigma$. 

The take home message from this section is that anyons are charge-flux particles in gauge theories and energy-flux particles in gravity.

\section{The Classical Anyon Impulse}
One way to understand the classical double copy is to study the classical impulse, a useful observable that can be computed using both the equations of motion and directly from the scattering amplitudes \cite{Maybee:2019jus}. In this section, we will compute the impulse imparted to a light probe particle in the background of a heavy, charged, spinning anyon as well as its gravitational counterpart. This mirrors the situation in 3+1 dimensions where we might consider a spinning dyon background, whose gravitational counterpart is Kerr-Taub-NUT \cite{Emond:2020lwi}.
\subsection{Electromagnetic Scalar Impulse}\label{sec:emscalar}
As a warmup, and for comparison later on, we will begin by calculating the impulse imparted to a scalar particle probing an electromagnetically charged heavy scalar, governed by the action in eq. \eqref{EMaction}, but with $k=0$, meaning we are considering a massless mediating particle.

At zeroth order in the perturbative expansion, we can assume a straight-line trajectory for the test particle, which we parametrise as $x_1^\mu= b^\mu + u_1^\mu\tau$, where $b^\mu$ is the impact parameter. The leading order impulse on particle 1 due to particle 2 is then given by
\begin{equation}
	\Delta p_1^\mu = g_1\int \sd\tau F_2^{\mu\nu}(x)u_{1\nu} \;.
\end{equation}
We will consider a static point source at the origin, with current
\begin{equation}
	J^\mu_2 = -g_2u_2^\mu\delta^{(2)}(\mathbf{r}) \;,
\end{equation}
where $g$ is the coupling and $u_{2\mu} = (1,0,0)$. In this case, the equations of motion are given by
\begin{equation}
	\pd_\mu F^{\mu\nu} = -g_2u_2^\nu\delta^{(2)}(\mathbf{r}) \;,~~~~~\pd_\mu \tilde{F}^\mu = 0 \;,
\end{equation}
where the dual tensor is given by $\tilde{F}^\mu = \frac12\epsilon^{\mu\nu\rho}F_{\nu\rho}$. We can rewrite this in momentum space as
\begin{equation}
	iq_\mu F^{\mu\nu} = iq_\mu \epsilon^{\mu\nu\rho}\tilde{F}_\rho = -g_2u_2^\nu\delta(q^0) \;.
\end{equation}
Acting on both sides with $\epsilon_{\nu\alpha\beta} q^\beta$, we then find
\begin{equation}
	i\epsilon_{\nu\alpha\beta}q^\beta\epsilon^{\mu\nu\rho}q_\mu\tilde{F}_\rho = iq^2 \tilde{F}_\alpha -iq_\A (q\cdot \tilde{F}) = iq^2 \epsilon_{\nu\alpha\beta}F^{\nu\beta} = g_2\epsilon_{\nu\alpha\beta}u_2^\nu q^\beta\delta(q^0) \;.
\end{equation}
Isolating the tensor and rewriting this covariantly, we then find that
\begin{equation}
	F_2^{\mu\nu}(x) = ig_2\int \sd^3\hat{q}\,e^{iq\cdot x}~\frac{\delta(q\cdot u_2)u_2^{[\mu}q^{\nu]}}{q^2}\;.
\end{equation} 
Plugging this into the impulse formula, we find
\begin{align}\label{eq:pointchargeimpulse}
	\Delta p_1^\mu &= ig_1g_2\int \sd\tau \int \sd^3\hat{q}\,e^{iq\cdot x} \delta(q\cdot u_2)\frac{u_2^{[\mu}q^{\nu]}u_{1\nu}}{q^2}\nn\\
	&= -ig_1g_2 \int \sd^3\hat{q}\,e^{iq\cdot b}\delta(q\cdot u_1)\delta(q\cdot u_2) \frac{q^\mu \cosh w}{q^2} \;, 
\end{align}
where we have used $u_1\cdot u_2 = -\cosh w$ in which $w$ is the rapidity.

The integrand is identical to that in four dimensions, and we note that this can in fact be derived using spinors only by considering the kick suffered by one spinor using the spinorial equations of motion, as was recently considered in \cite{Guevara:2020xjx}.  
\subsection{Electromagnetic Anyon Impulse}\label{sec:emanyon}
We will now investigate what happens if we introduce a Chern-Simons term into the mix (i.e. $k\neq 0$ above), rendering the photon topologically massive and turning the matter particles into anyons.

Since anyons have spin, we will consider a cylindrically symmetric rotating point source described by a current
\begin{equation}
	J^\mu_2 = -\left(g_2u_2^\mu + \tilde{g}_2\epsilon^{\mu\nu\rho}u_{2\nu}\pd_\rho\right)\delta^{(2)}(\mathbf{r})\;,
\end{equation}
where $g_2$ and $\tilde{g}_2$ are couplings and $u_{2\mu} = (1,0,0)$. 

Including this source, the equations of motion are given by
\begin{equation}
	\left(\eta_{\mu\rho}\pd_\nu + \frac{ke^2}{2}\epsilon_{\mu\nu\rho}\right)F^{\nu\rho} = -\left(g_2u_2^\mu + \tilde{g}_2\epsilon^{\mu\nu\rho}u_{2\nu}\pd_\rho\right)\delta^{(2)}(\mathbf{r})\;,
\end{equation}
which we can rewrite in terms of the (Hodge) dual vector field $\tilde{F}^{\mu} \equiv \frac12\epsilon^{\mu\nu\rho}F_{\nu\rho}$ as
\begin{equation}
	\left(m\eta_{\mu\nu} + \epsilon_{\mu\nu\rho}\pd^\rho\right)\tilde{F}^{\nu} = \cl{G}_{\mu\nu}\tilde{F}^{\nu} = -\left(g_2u_2^\mu + \tilde{g}_2\epsilon^{\mu\nu\rho}u_{2\nu}\pd_\rho\right)\delta^{(2)}(\mathbf{r})\;,
\end{equation}
where we have identified $m=ke^2$ as the mass of the photon. Then, using the fact that $\cl{G}^{\mu\alpha}\cl{G}_{\alpha\nu}\tilde{F}^\nu = (\pd^2 - m^2)\tilde{F}^\mu + 2m \cl{G}^{\mu}_{~\nu} \tilde{F}^\nu$, we can write
\begin{equation}
	(\pd^2 - m^2)\tilde{F}^\mu = (\cl{G}^{\mu\nu} -2m\eta^{\mu\nu})J_{2\nu}\;.
\end{equation}
Expressing the current in momentum space, taking $x_2 = b + u_2\tau$, we find
\begin{align}
	J_{2}^\mu(q) &= -\left(g_2u_2^\mu + i\tilde{g}_2\epsilon^{\mu\nu\rho}u_{2\nu}q_\rho\right)\int \sd^3x e^{iq\cdot x}\delta^{(2)}(\textbf{x}-\textbf{x}_2) \nn\\
	&= -e^{iq\cdot b}\delta(q\cdot u_2)\left(g_2u_2^\mu + i\tilde{g}_2\epsilon^{\mu\nu\rho}u_{2\nu}q_\rho\right)\;.
\end{align}
Using this, we can write the momentum space dual tensor as
\begin{equation}
	\tilde{F}^\mu = (g_2+\tilde{g}_2 m)\frac{\delta(q\cdot u_2)e^{-iq\cdot b}}{q^2 + m^2}(mu_2^\alpha - i\epsilon^{\alpha\beta\mu}q_\beta u_{2\mu}) \;,
\end{equation}
such that
\begin{equation}
	F_{\mu\nu}(q) = \epsilon_{\mu\nu\rho}\tilde{F}^\rho = (g_2+\tilde{g}_2 m)\frac{\delta(q\cdot u_2)}{q^2 + m^2}(m\epsilon_{\mu\nu\rho}u_2^\rho - 2iq_{[\mu} u_{2\nu]})\;.
\end{equation}
Plugging this in to get the impulse then gives
\begin{align}\label{eq:gauge impulse}
	\Delta p_1^\mu &= g_1\int \sd\tau F_2^{\mu\nu}(x)u_{1\nu} = ig_1(g_2+\tilde{g}_2 m)\int \sd\tau\int \hat{\sd}^3q\,e^{-iq\cdot x}\frac{\delta(q\cdot u_2)}{q^2 + m^2}\left(q^\mu (u_1\cdot u_2) - im\epsilon^{\mu\nu\rho}u_{1\rho} u_{2\nu}\right)\nn\\
	&= ig_1(g_2+\tilde{g}_2m)\int \hat{\sd}^3q\,\delta(q\cdot u_1)\delta(q\cdot u_2)e^{-iq\cdot x}\frac{\left(q^\mu \cosh w - im\epsilon^{\mu\nu\rho}u_{1\rho} u_{2\nu}\right)}{q^2 + m^2} \;.
\end{align}
This shares a similarity with the massless case, now introducing a dimension-specific Levi-- Civita component to the integrand.
\subsection{Gravitational Scalar Impulse}\label{sec:conical}
Pure Einstein gravity is topological in $2+1$ dimensions, and there are no propagating gravitons. However, when coupled to matter there are long range effects, which vanish in the non--relativistic (NR) limit \cite{Deser:1983tn,Marcus:1983hb}. Since there are no propagating degrees of freedom, the interaction is purely topological, originating from the conical singularity created by a matter source at the origin \cite{deSousaGerbert:1988qzd}. Surprisingly, this does generate a non-zero scattering amplitude and thus will give rise to a non-zero impulse, albeit one that vanishes in the NR limit. Before seeing whether or not such a solution can be derived from purely on-shell techniques, let's look at the calculation of the impulse using the standard formulation.

The spacetime we will consider is flat everywhere except at $r=0$, which has a conical singularity generated by a point mass at the origin, with Ricci scalar
\begin{equation}\label{ricciconical}
	R = 8\pi GM\delta^{2}(\mathbf{r}) \;.
\end{equation} 
This is the natural analogue of the point charge discussed in section \ref{sec:emscalar}, and therefore we might naturally expect it to show up in the double copy somehow. The solution of Einstein's equation is given by the line element
\begin{equation}
	\sd s^2 = -\sd t^2 + r^{-8GM}(\sd r^2 + r^2\sd\phi^2) \;,
\end{equation}
which is indeed singular only at the origin. Treating this perturbatively in $G$, we can express the deviation from Minkowski as
\begin{equation}
	h_{\mu\nu} = -8GM\log r\left(\eta_{\mu\nu} + u_\mu u_\nu\right) = 2\kappa^2M\left(\eta_{\mu\nu} + u_\mu u_\nu\right)C(r) \;,
\end{equation}
where for later convenience we define the function $C(r) = -\frac{1}{2\pi}\log(r)$ and $u_\mu = (1,0,0)$.

We can derive the geodesic of a particle with momentum $p_1^\mu$ moving in the spacetime generated by particle 2, at linear order, using the spin-connection via
\begin{equation}
	\frac{\sd u^\mu_1}{\sd\tau} = -\omega^\mu_{\nu\rho}u_1^\nu u_1^\rho + \pd_\rho h^\mu_{~\nu}u_1^\nu u_1^\rho = -\omega^\mu_{\nu\rho}u_1^\nu u_1^\rho +  \frac{\pd}{\pd\tau}\left(h^{\mu\nu}u_{1\nu}\right) \;,
\end{equation}
where we can express $\omega_{\mu\nu\rho} = \pd_{[\nu} h_{\nu]\rho}$. We will show explicitly in the next section that the surface term can be discarded, but for now it will be assumed.

Working in momentum space, we define the Fourier-dual of the spin-connection as
\begin{align}
	\Omega_{\mu\nu\rho}(q) = \int \sd^3x\,e^{-iq\cdot x}\,\pd_{[\nu} h_{\nu]\rho}
	\ = \ i\hat{\delta}(u\cdot q)q_{[\mu}h_{\nu]\rho}(q) \;.
\end{align}
Making use of the fact that the Fourier transform is given by $\cl{F}[C(r)] = \frac{1}{q^2}$, we find that the momentum space metric generated by a point particle with mass $m_2$ is given by
\begin{equation}
	h_{\mu\nu}(q) = 2\frac{\kappa^2m_2}{q^2}\left(\eta_{\mu\nu} + u_{2\mu} u_{2\nu}\right) \;. 
\end{equation}
To evaluate the impulse imparted to particle one (at leading order), we need to integrate the geodesic, i.e.
\begin{equation}
	\Delta p_1^\mu = m_1\int \sd\tau \frac{\sd u^\mu_1}{\sd\tau} = m_1\int \sd\tau\int \hat{\sd}^3q~ e^{iq\cdot (b + u_1\tau)} \Omega^\mu_{~\nu\rho}u_1^\nu u_1^\rho\nn \;.
\end{equation}
Plugging everything in, we then find
\begin{align}\label{conicalimp}
	\Delta p_1^\mu &= i\kappa^2 m_1m_2\int \hat{\sd}^3q~\hat{\delta}(u_1\cdot q)\hat{\delta}(u_2\cdot q) e^{iq\cdot b} \frac{q^\mu}{q^2}\left((u_1\cdot u_2)^2-1\right) \nn \\
	&= i\kappa^2 m_1m_2\int \hat{\sd}^3q~\hat{\delta}(u_1\cdot q)\hat{\delta}(u_2\cdot q) e^{iq\cdot b} \frac{q^\mu\sinh^2w}{q^2} \;.
\end{align}
This reassuringly vanishes in the NR $w\rightarrow 0$ limit as expected, meaning we have not discovered an unwelcome Newtonian potential. What is less reassuring is the fact that this doesn't seem to be the double copy of what we found section \ref{sec:emscalar}: the single copy is perfectly well defined in the $w\rightarrow 0$ limit. This is because the double copy can also include contributions from the dilaton, whereas this is only the (purely topological) graviton contribution. Adding the dilaton contribution then gives
\begin{align}\label{conicalimp}
	\Delta p_1^\mu &= i\kappa^2 m_1m_2\int \hat{\sd}^3q~\hat{\delta}(u_1\cdot q)\hat{\delta}(u_2\cdot q) e^{iq\cdot b} \frac{q^\mu(\sinh^2w + 1)}{q^2} \nn\\
	&= i\kappa^2 m_1m_2\int \hat{\sd}^3q~\hat{\delta}(u_1\cdot q)\hat{\delta}(u_2\cdot q) e^{iq\cdot b} \frac{q^\mu\cosh^2w}{q^2} \;.
\end{align}
\textcolor{red}{Referring back to the electromagnetic impulse given in eq.~\eqref{eq:pointchargeimpulse},} we find then that the double copy of the massless photon is a combination of the topological graviton and the dilaton. This is explored in more detail in appendix \ref{app:sugra}.
\subsection{Gravitational Anyon Impulse}\label{sec:granyon}
Massive gravitons in $2+1$ dimensions have degrees of freedom and can indeed propagate locally. Adding a gravitational Chern-Simons term to the Einstein action converts coupled particles to \textit{gravitational anyons} \cite{Deser:1989ri,Ortiz:1991gx}, in exact analogy to the electromagnetic case. We will compute the impulse of a particle probing a gravitational anyon, considering topologically massive gravity \cite{Deser:1981wh} coupled to matter, given by the action
\begin{equation}
	S_{TMG} = -\frac{1}{2\kappa^2}\int \sd^3x\sqrt{-g}\left[R + \frac{1}{2m}\epsilon^{\lambda\mu\nu}\Gamma^\rho_{\lambda\sigma}\left(\pd_\mu\Gamma^\sigma_{\nu\rho} + \frac{2}{3}\Gamma^\sigma_{\mu\tau}\Gamma^\tau_{\nu\rho}\right)\right] + S_{matter}[\phi;g_{\mu\nu}] \;,
\end{equation}
where $m$ is the topological mass parameter, and $\kappa^2=8\pi G$. One important point about this action is that the sign of the Einstein part of the action is opposite to the usual convention. This is to ensure that the physical mode is not ghost like \cite{Deser:1981wh}. Moreover, despite the theory being massive, it nevertheless remains diffeomorphism invariant (without needing to resort to the St\"{u}ckelburg trick). This is a result of the Chern-Simons term transforming as a total derivative under a diffeomorphism \cite{Deser:1981wh}. We will consider the same matter action as in the gauge theory case, where now the gauge-covariant derivative is simply replaced with the minimally coupled kinetic term.

The equations of motion, noting the minus sign on the stress-energy tensor, are given by
\begin{equation}\label{eq:TMG eom}
	G_{\mu\nu} + \frac{1}{m}C_{\mu\nu} \ = \ -\kappa^2 T_{\mu\nu} \;,
\end{equation} 
where $G_{\mu\nu} \equiv R_{\mu\nu} - \frac12Rg_{\mu\nu}$ is the usual Einstein tensor and $C_{\mu\nu} \equiv \epsilon_\mu^{~\rho\sigma}\nabla_\rho\left(R_{\sigma\nu} - \frac14 g_{\sigma\nu}R\right)$ is the Cotton tensor, the three dimensional equivalent of the Weyl tensor. We note that Einstein's equations can be recovered by taking $m\rightarrow\infty$ and by a rescaling of $\kappa$, but that the $m\rightarrow 0$ limit is not so well defined. We will return to this point later.

Expanding eq. \eqref{eq:TMG eom} around a Minkowski background $g_{\mu\nu}=\eta_{\mu\nu}+h_{\mu\nu}$, and working in the de Donder gauge $\pd^\mu h_{\mu\nu} = \frac12\pd_\nu h$, the linearized equations of motion are
\begin{equation}
	\pd^2h_{\mu\nu} - \frac12\eta_{\mu\nu}\pd^2h + \frac{1}{m}\pd_\rho\epsilon^{\rho\lambda}_{\;\;\;(\mu}\pd^2 h_{\nu)\lambda} = 2\kappa^2 T_{\mu\nu} \;.
\end{equation}
For simplicity, we can express this in a trace reversed form as
\begin{equation}\label{eq:operator h eom}
	 \mathcal{O}_{\mu\nu\rho\lambda}(m)\,\pd^2h^{\rho\lambda} = 2\kappa^2(T_{\mu\nu} - \eta_{\mu\nu}T) \;,
\end{equation}
where we defined $\mathcal{O}_{\mu\nu\rho\lambda}(m) \equiv \mathcal{I}_{\mu\nu\rho\lambda} + \frac{1}{m}\pd_\alpha\epsilon^{\alpha\beta}_{\;\;\;(\mu}\mathcal{I}_{\nu)\beta\rho\lambda}$, with $\mathcal{I}_{\mu\nu\rho\lambda}\equiv\frac{1}{2}\big(\eta_{\mu\lambda}\eta_{\nu\rho}+\eta_{\mu\rho}\eta_{\nu\lambda}\big)$.

By operating from the left on eq. \eqref{eq:operator h eom} with $\mathcal{O}_{\mu\nu\rho\lambda}(-m)$, we can recast the linearized equations of motion as 
\begin{equation}
	\left(\pd^2 - m^2\right)\pd^2 h_{\mu\nu} = -2\kappa^2 m^2\left[T_{\mu\nu} - \eta_{\mu\nu}T - \frac{1}{m}\pd_\rho\epsilon^{\rho\lambda}_{\;\;\;(\mu}T_{\nu)\lambda} + \frac{1}{2m^2}\left(\eta_{\mu\nu}\pd^2 + \pd_\mu\pd_\nu\right)T\right] \;.
\end{equation}
Note that in the vacuum case ($T_{\mu\nu}=0$), the field equations reduce to 
\begin{equation}
	\left(\pd^2 - m^2\right)\pd^2 h_{\mu\nu} = 0\;,
\end{equation}
implying that we have found a propagating particle with mass $m$.

Let us now consider the case of a time-independent circularly symmetric matter source. This can be treated as a spinning point mass, which we take to be at the spatial origin of our coordinate system. The corresponding stress-energy tensor is then given by
\begin{equation}\label{eq:matter source}
	T_{\mu\nu} \ = \ \left(Mu_\mu u_\nu + \sigma u_{(\mu}\epsilon_{\nu)}^{~~\A\B}u_\A\pd_\B\right)\delta^{(2)}(\mathbf{r}) \;,
\end{equation}
where $\sigma$ is the spin and $u_\mu = (1,0,0)$.

In the presence of a time-independent circularly symmetric matter source \eqref{eq:matter source}, we can write the metric (in the de Donder gauge) as\footnote{Note that our expression for $h_{\mu\nu}$ differs somewhat from that in refs. \cite{Deser:1989ri,Ortiz:1991gx}, as the authors work in a conformally spatial gauge, as opposed to the de Donder gauge. This is not a surprise, as $h_{\mu\nu}$ is a gauge-dependent object and not an observable. Of course, all observable quantities will ultimately be independent of the gauge we choose.}
\begin{flalign}
	h_{\mu\nu}(r) =& \ \kappa^2(M + m\sigma)\big[\eta_{\mu\nu} + 2u_\mu u_\nu\big]Y(r) - 2\kappa^2 M\big[\eta_{\mu\nu} + u_\mu u_\nu\big]C(r) \nn \\ & \ - \frac{\kappa^2}{m^2}(M + m\sigma)\big[2m u_{(\mu}\epsilon_{\nu)}^{\;\;\rho\lambda}u_\rho\pd_\lambda - \pd_\mu\pd_\nu\big]\big[C(r) - Y(r)\big] \;,
\end{flalign}
where $Y(r) = \frac{1}{2\pi} K_0(mr)$ and $C(r) = -\frac{1}{2\pi}\ln r$, with $K_0(z)$ a modified Bessel function of the second kind, and we note that
\begin{equation}
	(\pd^2 - m^2)Y(r) \ = \ -\delta^{(2)}(\mathbf{r}) \;,\qquad \pd^2C(r) \ = \ -\delta^{(2)}(\mathbf{r}) \;.
\end{equation} 
Let us briefly examine the total derivative term in the geodesic equation above. In this analysis, we are probing the spacetime generated by a heavy particle of mass $m_2$, with a much lighter particle of mass $m_1$. At leading order, the geodesic of the probe particle is a straight line, given by 
\begin{equation}
x^\mu_1(\tau) \ = \ b^\mu + u_1^\mu\tau \;.
\end{equation} 
Evaluating the total derivative term on this solution, we find that
\begin{flalign}
	\frac{\pd}{\pd\tau}\left(h^{\mu\nu}u_{1\nu}\right) \ =& \ -\frac{\kappa^2m(\mathbf{u}_1\cdot\mathbf{r})}{2\pi r}(m_2+m\sigma)\bigg[(u_1^\mu - 2\cosh w\,u_2^\mu)\,K_1(mr) \nn\\[0.4em] &\ +\frac{1}{m^2}\Big(2mu_2^{(\mu}\epsilon^{\nu)\rho\lambda}u_{1\nu}u_{2\rho}\pd_\lambda - (u_1\cdot\pd)\pd^\mu\Big)\left(K_1(mr)-\frac{1}{mr}\right)\bigg]\nn\\[0.4em]&\ + \frac{\kappa^2 m_2(\mathbf{u}_1\cdot\mathbf{r})}{\pi r^2}(u_1^\mu - \cosh w\,u_2^\mu) \;,
\end{flalign}
where we have used that $r^i=x^i$, with  $r=\sqrt{x^ix_i}$, such that $\frac{\partial r}{\partial\tau}=\frac{u_1^i x_i}{r}=\frac{\mathbf{u}_1\cdot\mathbf{r}}{r}$ (note that bold vectors are two-vectors, spatial components of the corresponding spacetime three-vectors), and that $u_1\cdot u_2 = -\cosh w$.

In order to neglect this total derivative contribution, we need it to fall off sufficiently quickly as $r\rightarrow\infty$, under the integral. We see that this is indeed the case, as $K_1(mr)\rightarrow 0$ (and clearly $\frac{1}{r}\rightarrow 0$) in this limit. We are therefore safe to ignore the total derivative term in this scenario.

Working in momentum space, we define the Fourier-dual of the spin-connection as
\begin{align}
	\Omega_{\mu\nu\rho}(q) = \int \sd^3x\,e^{-iq\cdot x}\,\pd_{[\mu} h_{\nu]\rho}
	\ = \ i\hat{\delta}(u\cdot q)q_{[\mu}h_{\nu]\rho}(q) \;.
\end{align}

We find then that the momentum space metric is given by
\begin{flalign}
	h_{\mu\nu}(q) =& \ \kappa^2\frac{(m_2 + m\sigma)}{q^2(q^2+m^2)}\left[q^2\left(\eta_{\mu\nu}+2u_{2\mu} u_{2\nu}\right) + 2iu_{2(\mu}\epsilon_{\nu)}(u_{2},q_\lambda) - q_\mu q_\nu\right] \nn\\[0.4em] &\ - 2\frac{\kappa^2 m_2}{q^2}\left(\eta_{\mu\nu}+u_{2\mu} u_{2\nu}\right) \;,
\end{flalign}
where we adopt the compact notation $\epsilon_\mu(a,b)=\epsilon_{\mu\nu\lambda}a^\nu b^\lambda$, $\epsilon(a,b,c)=\epsilon_{\mu\nu\lambda}a^\mu b^\nu c^\lambda$. In this derivation we have used the following Fourier transforms
\begin{equation}
	\cl{F}[Y(r)] = \frac{1}{q^2+m^2},~~~~~\cl{F}[C(r)] = \frac{1}{q^2},~~~~~\cl{F}[Y(r) - C(r)] = -\frac{m^2}{q^2(q^2+m^2)} \;,
\end{equation} 
where our conventions are detailed in appendix~\ref{sec:FT}.

We note the existence of the massless $q^{-2}$ pole that is not on-shell. This is what we refer to earlier as the massless spin-2 ghost, since while it exists it is purely topological and carries no degrees of freedom. However, if we wish to recover the correct massless limit (i.e. scattering in topological Einstein gravity), then we should expect such a pole to show up. This pole is similar in spirit to the spurious pole found when considering e.g. charge-monopole scattering in four dimensions \cite{Huang:2019cja,Moynihan:2020gxj,Terning:2020dzg,Emond:2020lwi}, which should drop out of any observable. However, as we will see, great care must be taken when considering such poles in 2+1 dimensions.

Plugging in the metric, we then find

\begin{flalign}
	\Omega_{\mu\nu\rho}(q) =& \ i\hat{\delta}(u_2\cdot q)\kappa^2\frac{(m_2 + m\sigma)}{q^2(q^2+m^2)}\Big[q^2\left(q_{[\mu}\eta_{\nu]\rho}+2q_{[\mu}u_{2\nu]} u_{2\rho}\right) + iq_{[\mu}u_{2\nu]}\epsilon_{\rho}(u_{2},q) \nn\\[0.8em] & \ + iu_{2\rho}q_{[\mu}\epsilon_{\nu]}(u_{2},q)\Big]  - 2i\hat{\delta}(u_2\cdot q)\frac{\kappa^2 m_2}{q^2}\left(q_{[\mu}\eta_{\nu]\rho}+q_{[\mu}u_{2\nu]} u_{2\rho}\right) \;.
\end{flalign}
At leading order, the impulse is then given by
\begin{align}
	\Delta p_1^\mu &= m_1\int \sd\tau \frac{\sd u^\mu_1}{\sd\tau} = m_1\int \sd\tau\int \hat{\sd}^3q~ e^{iq\cdot (b + u_1\tau)} \Omega^\mu_{~\nu\rho}u_1^\nu u_1^\rho\nn\\
	&= i\frac{\kappa^2 m_1}{2}\int \hat{\sd}^3q~ e^{iq\cdot  b}\hat{\delta}(u_1\cdot q)\hat{\delta}(u_2\cdot q)\,q^\mu\,\Bigg[\frac{(m_2+m\sigma)\left(2(u_1\cdot u_2)^2 - 1\right)}{q^2 + m^2}\nn\\
	&~~~~~~~~~~~~~~~~~~+ \frac{2im(m_2+m\sigma)(u_1\cdot u_2)\epsilon(u_1,u_2,q)}{q^2(q^2 + m^2)} - \frac{2 m_2\left((u_1\cdot u_2)^2 - 1\right)}{q^2}\Bigg] \;.
\end{align}
Using $\cosh w = -u_1\cdot u_2$, we can write this as
\begin{align}
	\Delta p_1^\mu &= i\frac{\kappa^2 m_1}{2}\int \hat{\sd}^3q~ e^{iq\cdot b}\hat{\delta}(u_1\cdot q)\hat{\delta}(u_2\cdot q) \nn\\&~~~~~~~~~~\times\Bigg[\frac{q^\mu(m_2+m\sigma)\cosh2w}{q^2 + m^2}
	- \frac{2im(m_2+m\sigma)\cosh w\epsilon^\mu(u_1,u_2)}{q^2 + m^2} - \frac{2q^\mu m_2\sinh^2 w}{q^2}\Bigg] \;,
\end{align}
where we have used $q^\mu\epsilon(u_1,u_2,q) = q^2\epsilon^\mu(u_1,u_2)$ and we note that the massless pole remains in the impulse. Classically, this is not a problem since we know such a pole corresponds to a massless spin-2 particle which carries no degrees of freedom. However, as we will soon see, this pole is crucial in obtaining the correct massless limit. In order to compare with our previously derived results, we can take the infinite mass limit (or large $q$) limit, finding
\begin{equation}
	\Delta p_1^\mu\bigg|_{m\rightarrow\infty} = -i\kappa^2 m_1m_2\int \hat{\sd}^3q~\hat{\delta}(u_1\cdot q)\hat{\delta}(u_2\cdot q) e^{iq\cdot b} \frac{q^\mu\sinh^2w}{q^2} \;.
\end{equation} 
Up to a sign, this is equal to the result we got from Einstein gravity, i.e. the conical singularity impulse of eq. \eqref{conicalimp}, with the sign difference coming from the fact that our sign convention was reversed in the topologically massive case. What is more surprising is that the zero-mass limit is also well defined here, giving rise to a pure-dilaton impulse of the form
\begin{align}
	\Delta p_1^\mu\bigg|_{m\rightarrow 0} &= i\kappa^2 m_1m_2\int \hat{\sd}^3q~\hat{\delta}(u_1\cdot q)\hat{\delta}(u_2\cdot q) e^{iq\cdot b} \frac{q^\mu(\cosh 2w-2\sinh^2w)}{q^2} \nn\\
	&= i\kappa^2 m_1m_2\int \hat{\sd}^3q~\hat{\delta}(u_1\cdot q)\hat{\delta}(u_2\cdot q) e^{iq\cdot b} \frac{q^\mu}{q^2} \;,
\end{align} 
where we note that the spin contribution has also dropped out, a manifestation of the fact that all massless particles are scalars (or fermions) in 2+1 dimensions. From the perspective of the action, it is peculiar that we might take the massless limit, given that it appears that the action diverges in this limit. However, this is perfectly natural from the perspective of an on-shell scattering amplitude. The fact that the limit of TMG has a dilaton-like limit when coupled to matter is a reflection of the massless non-dynamical `ghost' mode in the spectrum. This oddity is explored more in appendix \ref{app:dilaton}. 

We have derived both the classical electromagnetic anyon impulse and the gravitational anyon impulse from their respective actions hoping to see the emergence of some kind of classical double copy structure. It is not at all obvious that the gravitational anyon impulse is the double copy of the electromagnetic anyon impulse, and to understand the subtleties of this we now move on to compute the same quantities directly on-shell via scattering amplitudes, where we hope the double copy structure will be more manifest.
\section{Scattering Amplitudes}
\subsection{Spin Deformations in Three Dimensions}\label{sec:spindef}
In the previous section, we considered classically spinning bodies that produced an electro-- magnetic or gravitational field. In this section we wish to derive the same results directly from the on-shell scattering amplitudes, meaning we need to consider the scattering of spinning particles in the classical spin limit. In four dimensions, spin can be introduced by considering an exponential deformation of the scattering amplitude \cite{Guevara:2017csg,Guevara:2018wpp,Guevara:2019fsj,Huang:2019cja,Moynihan:2020gxj}, which can be interpreted as the on-shell equivalent of the Janis-Newman algorithm. Specifically, we can deform the relevant on-shell vertex by an exponential with exponent $a\cdot q$, where $a^\mu$ is the mass-rescaled Pauli-Lubanski pseudovector and $q_\mu$ is the transfer momentum, whose helicity information (little group weight) is carried by the $x$-factor at that vertex. In three dimensions, however, the spin is usually characterised by the Pauli-Lubanski pseudo\textit{scalar}, given by
\begin{equation}\label{pseudoscalar}
	\tilde{\sigma} = \epsilon^{\mu\nu\rho}\sigma_{\mu\nu}P_\rho = J\cdot P \;,
\end{equation}
where $\sigma_{\mu\nu} = \frac{1}{4}[\gamma_\mu,\gamma_\nu] = \frac12\epsilon_{\mu\nu\rho}\gamma^\rho$ is the angular momentum tensor and $P_\rho$ the three-momentum. \textcolor{red}{Note that we have suppressed the spinor indices here for brevity. In the following discussion we shall briefly make them explicit in the interest of clarity.}
 
Acting on a spin-$s$ wavefunction $\psi$, the pseudoscalar gives
\begin{equation}
	\tilde{\sigma}^{\A_1,\A_2\cdots \A_{s}}_{~~~\B_1\B_2\cdots \B_{2s}}\psi^{\B_1\B_2\cdots \B_{2s}} = -ms\,\delta^{\A_1,\A_2\cdots \A_{2s}}_{~~~\B_1\B_2\cdots \B_{2s}}\psi^{\B_1\B_2\cdots \B_{2s}} \;.
\end{equation}
For $s = \pm 1/2$, we find that $J^\mu = \frac12\gamma^\mu$ and this is none other than the Dirac equation
\begin{equation}
	(P_{\alpha\beta} \pm m\delta_{\A\B})\psi^{\beta} = 0 \;.
\end{equation}
While it is interesting to consider finite spin, we wish to examine a particle with classical angular momentum, meaning we consider a general spin--$s$ particle with momentum $p_2^\mu$ generating a classical field (spacetime), which we will probe using a charged (neutral) scalar, with the probe and background particles exchanging a topologically massive field with momen-- tum $q^\mu$. In order to formally take the classical spin limit, we need to consider building our amplitudes out of coherent spin states. In four dimensions, such states are simply SU(2) coherent states which, in the infinite spin limit, reduce to an ensemble of the minimum-- uncertainty coherent states of the harmonic oscillator. In three dimensions, the equivalent are $SU(1,1)$ coherent states | \textit{pseudo}spin coherent states | which have a large spin limit (often called the Bargmann limit) which reduces to the usual $2+1$ dimensional minimum--uncertainty states \cite{Bargmann:1946me}.

Using this definition of the spin, along with the group algebra, we find that we can define the spin vector of particle $i$ as
\begin{equation}
	a^\mu_i = \frac{p^\mu_i}{m_i} = u_i^\mu \;.
\end{equation} 

This replaces the four-dimensional (rescaled) Pauli-Lubanski pseudovector, and we can use it to deform our spinor helicity variables by an exponential factor, which in 4D is the equivalent of the Newman-Janis shift \cite{Guevara:2018wpp,Bautista:2019tdr,Guevara:2019fsj,Arkani-Hamed:2019ymq,Moynihan:2019bor,Burger:2019wkq,Emond:2020lwi}. We can therefore deform a scalar three-particle amplitude as
\begin{equation}
	\cl{A}_3 \rightarrow \cl{A}_3e^{\frac{a_i\cdot q}{m}} = \cl{A}_3e^{\sigma\frac{m}{m_i}} \;,
\end{equation}
where $m$ is the internal particle mass (i.e. $q^2 = -m^2$) and $\sigma$ is the classical spin. A full derivation of this formula is given in appendix \ref{app:spindef}.

\subsection{Electromagnetic Anyon Impulse from Amplitudes}
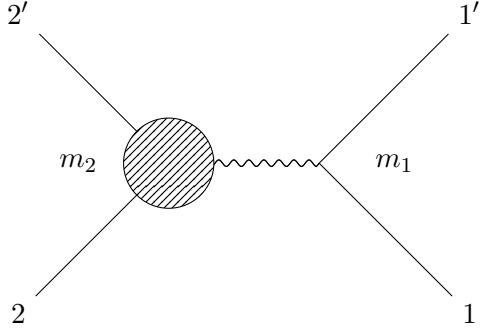
\begin{figure}[H]
	\centering
	\begin{tikzpicture}[scale=1]
	\begin{feynman}  
	\vertex (a) at (-4,2) {$2'$};
	\vertex (b) at (-4,-2) {$2$};
	\vertex (c) at (2,-2) {$1$};
	\vertex (d) at (2,2) {$1'$};
	\vertex (r) at (0,0);
	\vertex (l) at (-2,0) ;
	\diagram* {
		(a) -- [plain] (l) -- [photon] (r) -- [plain] (d),
		(b) -- [plain] (l) -- [photon] (r) -- [plain] (c),
	};
	\draw[preaction={fill, white},pattern=north east lines] (-2,0) ellipse (0.6cm and 0.6cm);
	%\draw[preaction={fill, white},pattern=north east lines] (-2,0) ellipse (0.6cm and 0.6cm);
	%=\draw[preaction={fill, white},pattern=north east lines] (0,0) ellipse (0.6cm and 0.6cm);
	%\draw [densely dashed, red, line width=0.5mm,] (0,1.2) -- (0,-1.6);
	\draw (-3.2,-0.25) node[above] {$m_2$};
	\draw (1,-0.25) node[above] {$m_1$};
	%\draw (-1,-0.65) node[above] {$\ell_2$};
	%\draw [densely dashed, red, line width=0.5mm,] (-1,0) -- (0,-1.6);
	\end{feynman}
	\end{tikzpicture}
	\caption{Electromagnetic probe of a charged classical anyon}
	\label{setup}
\end{figure}

In this section, we will calculate the leading order scattering amplitude of a light probe particle in the background of a static heavy charged anyon. We will consider the $2\rightarrow2$ scattering of a scalar of mass $m_1$ and \textcolor{red}{(electric) charge $e_1$}, and an anyon of mass $m_2$ and \textcolor{red}{(electric) charge $e_2$}, mediated by a massive gauge boson with mass $m$, as in Fig. \ref{setup}. The Mandelstam variables are defined as
\begin{equation}
s = -(p_1+p_2)^2,~~~~~t=-(p_1 + p_1')^2,~~~~~u = -(p_1+p_2')^2 \;.
\end{equation}

The scattering amplitude, for small $q^2 = -m^2$, is given by
\begin{align}
\cl{A}_4[1,2,1',2'] = \frac{\cl{A}_3[1,1',q^+]\cl{A}_3[2,2',q^-]}{q^2+m^2} \;,
\end{align}
where we have explicitly chosen the parity-violating \textit{positive} spin gauge boson\footnote{In principle, we could have started with a parity symmetric theory | where we would sum over the x--ratios | and performed a 3D duality rotation \cite{Huang:2019cja,Moynihan:2020gxj} to expose the parity--violating part. However, since we are interested in a parity--violating theory with distinct spin, this is not required and in fact the $x$-ratio we consider specifically picks out a helicity, with the inverse giving the opposite choice. From the perspective of the action, this corresponds to the choice of sign of the Chern--Simons term.}. 
In order to compute the on--shell three--particle amplitudes, we will use the formalism developed in \cite{Moynihan:2020ejh}, with the three--particle amplitudes given by
\begin{equation}
\cl{A}_3[1,1',q^+] = \sqrt{2}e_1m_1x_1,~~~~~\cl{A}_3[2^s,2'^s,q^{-}] = \sqrt{2}e_2m_2x_2^{-1}e^{\frac{m}{m_2}\sigma} \;.
\end{equation}
where $x_i = \frac{\braket{q|u_i|q}}{m}$. We note that the coupling constant must have mass dimension $[g] = 3/2 - s$ in order to be consistent.

We find that the amplitude in the small $q^2$ limit is given by
\begin{align}
\cl{A}_4[1,2,1',2'] &= \frac{2e_1e_2m_1m_2}{q^2+m^2}e^{\frac{m}{m_2}\sigma}\frac{x_1}{x_2} \nn \\
&= \frac{2e_1m_1m_2}{q^2+m^2}e^{\frac{m}{m_2}\sigma}\left(u_1\cdot u_2 + i\frac{m\epsilon(u_1,u_2,q)}{q^2} + \frac{m^2}{4m_1m_2}\right)\label{osamp} \;.
\end{align}

Given the scattering amplitudes, we can compute the impulse via
\begin{align}
	\Delta p_1^\mu &= \int \hat{\sd}^3q~\hat{\delta}(2p_1\cdot q + q^2)\hat{\delta}(2p_2\cdot q - q^2)\,e^{-iq\cdot b}iq^\mu \cl{A}_4[1,2,1',2'] \nn\\
	&= \frac{1}{4m_1m_2}\int \hat{\sd}^3q~\hat{\delta}(u_1\cdot q - \frac{m^2}{2m_1})\hat{\delta}(u_2\cdot q + \frac{m^2}{2m_2})\,e^{-iq\cdot b}iq^\mu \cl{A}_4[1,2,1',2'] \nn\\
	&\simeq \frac{1}{4m_1m_2}\int \hat{\sd}^3q~\hat{\delta}(u_1\cdot q)\hat{\delta}(u_2\cdot q)\,e^{-iq\cdot b}iq^\mu \cl{A}_4[1,2,1',2'] \;.
\end{align}
Expanding the amplitude to first order in $m$, we find
\begin{equation}
	\Delta p_1^\mu = \frac{e_1e_2}{2}\int \hat{\sd}^3q~\hat{\delta}(u_1\cdot q)\hat{\delta}(u_2\cdot q)\,e^{-iq\cdot b}iq^\mu\left(1+\frac{m}{m_2}\sigma\right)\left(\frac{(u_1\cdot u_2)}{q^2+m^2} + i\frac{m\epsilon(u_1,u_2,q)}{q^2(q^2+m^2)}\right) \;.
\end{equation}
Using the relationship
\begin{equation}
	q^\mu\epsilon(q,u_1,u_2) = -m^2\epsilon^\mu(u_1,u_2) + \cl{O}(m^2/m_1) \;,
\end{equation}
we find that the impulse simplifies to become
\begin{equation}
	\Delta p_1^\mu = \frac{ie_1e_2}{2}\int \hat{\sd}^3q~\hat{\delta}(u_1\cdot q)\hat{\delta}(u_2\cdot q)\,e^{-iq\cdot b}\left(1+\frac{m}{m_2}\sigma\right)\left(\frac{q^\mu(u_1\cdot u_2)}{q^2+m^2} - i\frac{m\epsilon^\mu(u_1,u_2)}{q^2+m^2}\right) \;.
\end{equation}
which matches the classical computation when we consider $g_i = e_i$ and $\tilde{g}_2 = e_2\frac{\sigma}{m_2}$ (cf. eq.~\eqref{eq:gauge impulse}).

\subsection{Gravitational Anyon Impulse from Amplitudes}
\begin{figure}[H]
	\centering
	\begin{tikzpicture}[scale=1]
		\begin{feynman}  
			\vertex (a) at (-4,2) {$2'$};
			\vertex (b) at (-4,-2) {$2$};
			\vertex (c) at (2,-2) {$1$};
			\vertex (d) at (2,2) {$1'$};
			\vertex (r) at (0,0);
			\vertex (l) at (-2,0) ;
			\diagram* {
				(a) -- [plain] (l) -- [graviton] (r) -- [plain] (d),
				(b) -- [plain] (l) -- [graviton] (r) -- [plain] (c),
			};
			\draw[preaction={fill, white},pattern=north east lines] (-2,0) ellipse (0.6cm and 0.6cm);
			%\draw[preaction={fill, white},pattern=north east lines] (-2,0) ellipse (0.6cm and 0.6cm);
			%=\draw[preaction={fill, white},pattern=north east lines] (0,0) ellipse (0.6cm and 0.6cm);
			%\draw [densely dashed, red, line width=0.5mm,] (0,1.2) -- (0,-1.6);
			\draw (-3.2,-0.25) node[above] {$m_2$};
			\draw (1,-0.25) node[above] {$m_1$};
			%\draw (-1,-0.65) node[above] {$\ell_2$};
			%\draw [densely dashed, red, line width=0.5mm,] (-1,0) -- (0,-1.6);
		\end{feynman}
	\end{tikzpicture}
	\caption{Gravitational probe of a charged classical anyon}
	\label{setup2}
\end{figure}
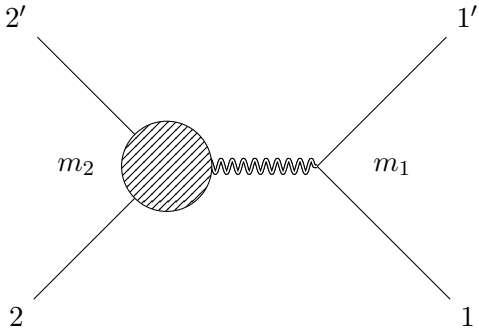

In this section, we will calculate the leading order scattering amplitude of a light probe particle in the background of a static heavy \textit{gravitational} anyon. We will consider the $2\rightarrow2$ scattering of a neutral scalar of mass $m_1$ and an anyon of mass $m_2$, mediated by a massive graviton with mass $m$, as in Fig. \ref{setup2}. 

Since we are considering a gravitational theory, the three-particle coupling is $g=\kappa_i$ by dimensional analysis, and we use the three particle amplitudes
\begin{equation}\label{grav3pts}
	\cl{M}_3[1,1',q^+] = \sqrt{2}\kappa m_1^2x_1^2,~~~~~\cl{M}_3[2^s,2'^s,q^{-}] = \sqrt{2}\kappa m_2^2x_2^{-2}e^{\frac{m}{m_2}\sigma} \;.
\end{equation}
In this form, the double copy is apparent since the prescription demands that we make the replacements $e_i\rightarrow\kappa m_i$ and $x \rightarrow x^2$.

The four-particle $t$-channel amplitude is then given by
\begin{align}
\cl{M}_4[1,2,1',2'] = \frac{\cl{M}_3[1,1',q^{+2}]\cl{M}_3[2^s,2{'s},q^{-2}]}{q^2+m^2} \;,
\end{align}
and plugging in the three-particle amplitude we get
\begin{align}
&\cl{M}_4[1,2,1',2']  =  2\kappa^2\frac{m_1^2 m_2^2}{q^2+m^2}e^{\frac{m}{m_2}\sigma}\left(\frac{x_1}{x_2}\right)^2 \nn\\[1em]
&\simeq 2\kappa^2\frac{m_1^2 m_2^2 }{q^2+m^2}e^{\frac{m}{m_2}\sigma}\left((u_1\cdot u_2)^2-\left( \frac{m\epsilon(u_1, u_2, q)}{q^2} \right)^2 +2iu_1\cdot u_2 \frac{m\epsilon(u_1, u_2, q)}{q^2} \right) \nn\\[1em]
&\simeq 2\kappa^2\frac{m_1^2 m_2^2 }{q^2+m^2}e^{\frac{m}{m_2}\sigma}\left(\cosh 2w + 2iu_1\cdot u_2 \frac{m\epsilon(u_1, u_2, q)}{q^2} \right) - 2\kappa^2e^{\frac{m}{m_2}\sigma}\frac{m_1^2 m_2^2 }{q^2}\sinh^2w \;, 
\end{align}
where we have used a partial fraction expansion to find
\begin{align}
	\frac{(u_1\cdot u_2)}{q^2+m^2}^2-\frac{m^2\epsilon(u_1, u_2, q)^2}{q^4(q^2+m^2)} = \frac{\cosh 2w}{q^2+m^2} - \frac{\sinh^2w}{q^2} \;.
\end{align}
Plugging this into the formula for the classical impulse, then, we find
\begin{align}
\Delta p_{1}^{\mu}
&= \frac{1}{4m_1m_2}\int \hat{\sd}^3q~\hat{\delta}(u_1\cdot q)\hat{\delta}(u_2\cdot q)\,e^{-iq\cdot b}iq^\mu \cl{M}_4[1,2,1',2']\nn\\
&= \frac{\kappa^2}{2}m_1\left(m_2 + m\sigma\right)\int \hat{\sd}^3q~\hat{\delta}(u_1\cdot q)\hat{\delta}(u_2\cdot q)\,e^{-iq\cdot b}\nn\\
&~~~~~~~~~~~~~~~~~~~~~~~~~~\times\left(\frac{q^\mu\cosh2w}{q^2+m^2}-\frac{2im\epsilon^\mu(u_1, u_2)\cosh w}{q^2+m^2} - \frac{q^\mu\sinh^2w}{q^2}\right) \;,
\end{align}
where we have again used $q^\mu\epsilon(u_1,u_2,q) = q^2\epsilon^\mu(u_1,u_2)$.
This does not \textit{precisely} match the impulse derived from topologically massive gravity via the metric, differing only by a factor of two on the last term. This factor is important, and it tells us a number of things. Firstly, it tells us that the there is an additional massless topological graviton contribution arising in the double copy, adding an extra term $\propto \sinh^2w$ when compared with the topologiclly massive case. This may seem peculiar, since at first glance it suggests that the double copy combines two vectors into \textit{two} spin--2 fields, one massive and one massless. This is not the case, however, since the massless field is a non--dynamical ghost, with no residue on--shell (at $q^2 = -m^2$) and no degrees of freedom (since it is purely topological). What is peculiar is the fact that there are twice the number in the double copy as there are in TMG. This peculiarity, a factor difference of two, is exactly what is required to ensure that the massless limit contains the correct mixing of the massless topological graviton and the massless dilaton, as found earlier in section  \ref{sec:conical} and which matches the classical double copy found in Ref. \cite{CarrilloGonzalez:2019gof}. Taking $m\rightarrow 0$ in the impulse, we find
\begin{align}
	\Delta p_{1}^{\mu}\bigg|_{m\rightarrow 0} &= \frac{\kappa^2}{2}m_1m_2\int \hat{\sd}^3q~\hat{\delta}(u_1\cdot q)\hat{\delta}(u_2\cdot q)\,e^{-iq\cdot b}\left(\frac{q^\mu\cosh2w - q^\mu\sinh^2w}{q^2}\right) \nn\\
	&= \frac{\kappa^2}{2}m_1m_2\int \hat{\sd}^3q~\hat{\delta}(u_1\cdot q)\hat{\delta}(u_2\cdot q)\,e^{-iq\cdot b}\left(\frac{q^\mu\cosh^2w}{q^2}\right) \;.
\end{align}
This is exactly what we would expect to get from double copying the point charge impulse\footnote{We note that this differs from the expected result by an overall factor of $1/2$, however this is due to the fact that we don't sum over helicities in this case, and would need to include the opposite helicity configuration (which would be identical) to obtain the correct factor. Alternatively we can simply rescale the coupling constant to match the result.} given by eq. \eqref{eq:pointchargeimpulse}, and we can easily see this by computing this directly from the massless $x$--ratio, which will give the same result. This mixing is not entirely surprising, since it is well known that the dimensional reduction of 4D General Relativity is a scalar--tensor theory \cite{Verbin:1994ks}, and in this case the extra scalar degree of freedom is added to give rise to the double copy. In the case of pure TMG, the massless limit is a dilaton, i.e.
\begin{equation}\label{coshn}
	\cosh2w - n\sinh^2w \ = \ \begin{cases}
		\cosh^2w&\quad n = 1,~~~\text{dilaton + graviton} \\[0.5em] 1\,&\quad n = 2,~~~\text{dilaton only.}
	\end{cases}
\end{equation}

In three dimensions, it is well known that the massless photon is dual to a massless scalar, since we can express $F^{\mu\nu} = \epsilon^{\mu\nu\rho}\pd_\rho\phi$, and so it is not unreasonable to expect that the massless photon might double copy to a massless dilaton. However, what is surprising is the contribution coming from a graviton, which as we saw in section \ref{sec:conical}, has no propagating degrees of freedom in 2+1 dimensions and is purely topological.

\section{Aharonov--Bohm Effect}
The Aharonov--Bohm (AB) phase is normally expressed in terms of a holonomy of the vector potential, e.g. the gauge potential produced by a solenoid perpendicular to the plane. Anyons are point particles with electric charge and a magnetic flux, where the magnetic flux is perpendicular to the plane and therefore the anyon behaves as an infinitely thin solenoid, which then gives rise to an AB phase. This phase can be found by taking a surface integral of the magnetic field produced by the anyon in the limit where all other interactions are disregarded. 

The phase itself is related to the non--relativistic (NR) potential energy $V(p,r)$ (that contains the vector potential), which we can relate to the amplitude via the Born approximation \textcolor{red}{in the center-of-mass frame:}
\begin{equation}\label{potentialexp}
	V(p,r) = \int\frac{\sd^{D-1}q}{(2\pi)^{D-1}}\,e^{i\mathbf{q}\cdot\mathbf{r}}\frac{1}{4E_{\mathbf{p}}E_{\mathbf{p}'}}\cl{M}(p,p') + \cdots \;,
\end{equation}
where \textcolor{red}{$\mathbf{q}=\mathbf{p}-\mathbf{p}'$ is the transfer three-momentum}, $(4E_{\mathbf{p}}E_{\mathbf{p}'})^{-1}\simeq (4m_1m_2)^{-1}$ takes into account the NR normalisation of external states, and the dots denote higher-order corrections that are suppressed in said limit. This relationship allows us to express the AB phase in terms of the on-shell scattering amplitude in the NR limit. The AB phase for non--relativistic systems is given by
\begin{equation}\label{abphase1}
	\alpha = \frac{ie}{\hbar}\oint_\gamma\frac{\textbf{p}\cdot \textbf{A}}{m} \sd t =\frac{ie}{\hbar}\oint_\gamma \textbf{A}\cdot \sd\textbf{r} \;,
\end{equation}
where we have taken $\textbf{p} = m\textbf{v} = m\frac{\sd\textbf{r}}{\sd t}$. Note that the particular path $\gamma$ of the line integral is not important, provided it encircles the anyon.  

At leading order, we can extract both the momentum space vector and scalar potentials of particle one from the scattering amplitude using eq. \eqref{potentialexp}, and then considering the NR potential felt by the probe particle (of mass $m_1$ and charge $e_1$) in the presence of an electric field generated by a massive source particle (of mass $m_2$ and charge $e_2$)
\begin{equation}
	V(p,q) = e_1\left(\frac{\textbf{p}_1\cdot \textbf{A}}{m_1} + \varphi(q)\right) = e_1\left(\textbf{v}_1\cdot \textbf{A} + \varphi(q)\right) \;,
\end{equation}
meaning we can express the potentials as
\begin{equation}
	\varphi(q) = \frac{1}{e_1}\frac{\cl{M}}{4m_1m_2}\bigg|_{\textbf{p}_1 = 0},~~~~~A_i(q) = \frac{1}{e_1}\frac{1}{4m_1m_2}\frac{\pd \cl{M}}{\pd v^i_1}\bigg|_{\textbf{v}_1=0} \;.
\end{equation}
In position space, the vector potential is therefore
\begin{equation}
	A_i(\textbf{r}) = \frac{1}{e_1}\frac{1}{4m_1m_2}\int \frac{\sd^{D-1}q}{(2\pi)^{D-1}}e^{i\textbf{q}\cdot \textbf{r}}\frac{\pd \cl{M}}{\pd v^i_1}\bigg|_{\textbf{v}_1=0} \;,
\end{equation}
leading to an expression of the AB phase in terms of the scattering amplitude
\begin{equation}\label{abphase}
	\alpha = \frac{i}{4\hbar\,m_1m_2}\oint_\gamma d\textbf{r}\int\frac{\sd^{D-1}q}{(2\pi)^{D-1}}e^{i\textbf{q}\cdot \textbf{r}}\frac{\pd \cl{M}}{\pd \textbf{v}_1}\bigg|_{\textbf{v}_1=0} \;.
\end{equation}
This expression is entirely general, however depending on the situation one must be careful to consider the appropriate limits to isolate the pure AB phase, as we will demonstrate in the next section.

\subsection{The Anyon Aharonov--Bohm Phase and its Double Copy}
The AB phase for anyons is a topological effect, arising from the pure Chern--Simons term in the action. However, we can treat this contribution as arising from topologically massive gauge theory in the limit where the gauge boson is infinitely heavy.
In our case, we need to consider $D = 3$ and $m\rightarrow \infty$, since we are working in three dimensions and we need to discard any contribution that might come from the Maxwell interactions: we need to consider the pure (topological) Chern--Simons limit. This is acheived by taking the $e\rightarrow \infty$ and $m \rightarrow\infty$ limits while keeping $e^2/m$ fixed \cite{Jackiw:1989nq, Deser:1990ve} and dropping all interaction terms. We consider then
\begin{align}
	\alpha &= \lim_{m\rightarrow\infty}\frac{i}{4\hbar\,m_1m_2}\oint_\gamma \sd\textbf{r}\int\frac{\sd^{2}q}{(2\pi)^{2}}e^{i\textbf{q}\cdot \textbf{r}}\frac{\pd \cl{M}}{\pd \textbf{v}_1}\bigg|_{\textbf{v}_1=0}\nn\\
	&= \lim_{m\rightarrow\infty}\frac{1}{4\hbar\,m_1m_2}\int\sd^2x\,\epsilon^{ij}\pd_i\int\frac{\sd^{2}q}{(2\pi)^{2}}e^{i\textbf{q}\cdot \textbf{r}}\frac{\pd \cl{M}}{\pd v^j_1}\bigg|_{\textbf{v}_1=0} \;.
\end{align}
The relevant amplitude in the anyon case we are interested in is 
\begin{equation}
	\cl{M}(p,q) = \frac{2e_1^2\,m_1m_2}{q^2+m^2}\,\frac{x_1}{x_2} \;.
\end{equation}
We then consider the derivative of this amplitude with respect to $u^i_1$. This is given by
\begin{align}
	\frac{\pd \cl{M}(p,q)}{\pd u^i_1}
	&= \frac{2e_1^2\,m_1m_2}{q^2+m^2}\frac{\pd}{\pd u_1^i}\left(\frac{x_1}{x_2}\right)\nn\\
	%	&= \frac{1}{4e_1m_2}\frac{\pd}{\pd p_1^i}\left(\frac{4e_1}{q^2+m^2}\left((p_1\cdot p_2 + \frac{m^2}{4}) - i\frac{m\epsilon(p_1,p_2,q)}{q^2}\right)\right)\\
	&= -\frac{2e_1^2\,m_1m_2}{q^2+m^2}\left(u_{2i} - (u_2\cdot q)\frac{q_i}{q^2} + i\frac{m\epsilon_i(u_2,q)}{q^2}\right) \;,
\end{align} 
Returning to position space, this gives 
\begin{equation}
	\frac{\pd \cl{M}(p,r)}{\pd v^i_1} = -\frac{e_1^2\,m_1m_2}{\pi}\left[u_{2i}\,K_0(mr) + \left(\frac{(u_2\cdot \pd)\pd_i}{m^2} - \frac{\epsilon_{i\mu\nu}u_2^\mu\pd^\nu}{m}\right)\left(K_0(mr) + \text{ln}(mr)\right)\right] \;,
\end{equation}
where we have made use of the Fourier transforms for $q^{-2}$ and $(q^2+m^2)^{-1}$, as defined in Appendix \ref{sec:FT}, and also noted that 
\begin{equation}
	\frac{1}{q^2(q^2+m^2)} = \frac{1}{m^2}\left(\frac{1}{q^2} - \frac{1}{q^2+m^2}\right) \;.
\end{equation}

Taking the Chern--Simons and NR limits \footnote{Note that in this limit $u_1^i\rightarrow v_1^i$.} then gives
\begin{equation}
	\frac{\pd \cl{M}(p,r)}{\pd v^i_1} = -\frac{e_1^2\,m_1m_2}{\pi m}\epsilon^{ij}\pd_j\log(r) = -\frac{e_1^2\,m_1m_2}{\pi m}\epsilon^{ij}\frac{r_j}{r^2} \;,
\end{equation}
such that the phase is then given by
\begin{equation}
	\alpha = \frac{1}{4\hbar\,m_1m_2}\oint \sd x^i\,\frac{\pd \cl{M}(p,r)}{\pd v^i_1}\bigg|_{\textbf{v}_1=0} = \frac{e_1^2}{4\pi m}\int~\sd^2x\,\epsilon^{ij}\,\pd_j\,\epsilon_{ik}\,\pd^k\log(r) = \frac{e_1^2}{2m} \;,
\end{equation}
where we have used $\epsilon^{ij}\pd_j\epsilon_{ik}\pd^k\log(r) = \pd^2\log(r) = 2\pi\delta^{(2)}(\mathbf{r})$. This is the celebrated Aharonov-Bohm phase for an anyon. Using the fact that $m = ke^2$, we can express this phase as
\begin{equation}\label{key}
	\alpha = \frac{1}{2k} = \frac{e\phi}{2}.
\end{equation}
This is related to the anyonic statistics we derived in section \ref{sec:anyons} via the spin--statistics relation of ``phase $= 2\pi \times$spin", where we find that indeed we can recover the intrinsic angular momentum of the anyon as first defined in eq. \eqref{anyonspin}.

The gravitational anyon also gives rise to a phase picked up due to a gravitational potential, which we can also calculate from the on-shell amplitude. We could plug the gravitational amplitude into eq. \eqref{abphase}, effectively squaring the $x$--ratios, in order to compute the phase of the gravitational anyon in the limit where $m\rightarrow \infty$ with $\kappa/m$ kept fixed. However, we have already shown that this amplitude double copies, and so we don't have to: we can simply use the double copy prescription of $e\rightarrow \kappa m_2$ to find that the celebrated gravitational anyon Aharonov-Bohm phase is given by
\begin{equation}
	\alpha = \frac{\kappa^2}{\hbar}\frac{m_2^2}{m} = \frac{8\pi G}{\hbar}\frac{m_2^2}{m} \;.
\end{equation}
This nicely agrees with the existing results in the literature \cite{Deser:1989ri,Ortiz:1991gx,Deser:1990ve} and is another neat example of the utility of the double copy.
\section{Discussion}
We have shown that many interesting properties of anyons in both gauge theory and gravity can be derived from a purely on-shell philosophy, utilising the properties of the little group, locality and unitarity to construct classical and quantum observables. Indeed, it is apparent from the examples presented in this paper, that the on-shell amplitudes story in the realm of $(2+1)$--dimensional Chern--Simons type gauge and gravity theories is far more nuanced and richer than one might naively expect, leading to some intriguing physical results.

In particular, we demonstrated that classical observables on the gauge theory side do indeed double copy, although in a rather unexpected manner, with the double copy mixing a massive graviton and a (topological) massless spin-2 ghost. The ghost mode is intriguing, since it gives rise to additional velocity-dependent topological terms in the scattering amplitude and classical impulse. While this may seem odd, it is not surprising: the massless spin-2 field in Einstein gravity gives rise to the same phenomena, and in the case of the double copy (and TMG) we see the same phenomena but with the `ghostly' sign which arises from the Ricci scalar term in the TMG action. What is unexpected is that the double copy contains two such ghost modes, whereas TMG contains only one. In the Einstein limit ($m\rightarrow \infty$), it is \textit{only} this ghost mode that contributes to the scattering amplitude, in the form of the usual conical singularity solution. Being topological, the observables in this limit are all contact-like --- and only for non-static sources -- giving rise to vanishing Newtonian potential. For static, conserved sources, there is no contribution from the ghost mode as expected, and as can be seen by taking the $w\rightarrow 0$ limit in the impulse.

When considering TMG, it might seem as if taking the massless limit is ill-defined, since it is the $m\rightarrow \infty$ limit that gives Einstein gravity. However, from the perspective of the on-shell amplitudes, this limit should exist, and in fact it is perfectly well defined. In the limit where $m\rightarrow 0$ the extra ghost mode conspires with the dilaton in a precise way to give rise to the classical 3D double copy originally discovered in \cite{CarrilloGonzalez:2019gof}, including the ghostly sign. This can be traced back to the fact that taking the residues for the different poles and taking the particles to be on-shell are not commuting operations, as was shown in \cite{Moynihan:2020ejh}. 

Indeed, this observation is intimately related to the discussion in appendix~\ref{app:dilaton}, where we consider a generic  propagator, parametrised by (dimensionless free parameter) $\xi$, containing both a massive graviton and massless ghost mode. The non-commutativity of the limits then leads to residues that precisely correspond to the different choices of $\xi$ that we explore. To elaborate, we find that by first taking the propagator fully on-shell, we recover precisely the required residues for the TMG double copy to hold. In this case, taking the massless limit leaves us with both a massless topological graviton and a massless dilaton in the spectrum. The solution mixes the two of them in precisely the manner expected from naively double copying the corresponding 3D electromagnetic theory. If one instead keeps the propagator initially off-shell, taking the residues first, then this no longer leads to what is expected of the double copy. Consequently, in the massless limit, one finds that there is no longer a massless graviton in the spectrum, only a massless dilaton, corresponding to the usual non-dynamical ghost mode. 

This provides a nice interpretation of the results derived in the main body, where there naively seemed to be disconnect between the massless limits of the standard classical field theory calculation, and that found using on-shell amplitude techniques. We see now, that for the results to match up, we need to be careful in determining in which order we take the limits, as interchanging the two corresponds to considering different massless theories.

It remains an open question as to which field theory leads to the amplitudes discovered via the double copy. Three dimensional gravity is an extremely rich subject \cite{Bergshoeff:2010ad}, and there are many possible candidates: new massive gravity \cite{Bergshoeff:2009hq,Bergshoeff:2009aq}, general massive gravity or bi-metric massive gravity \cite{Akhavan:2016hju} to name a few. All of these give rise to massive spin-2 gravitons, and in some limit they reduce to linearized TMG, and some contain additional modes, which could potentially contribute in the correct way. We leave this to future study -- perhaps the pure double copy, relating topologically massive Yang-Mills with topologically massive gravity \textit{without} matter, will shed some light on this.

In addition to exploring the double copy in 2+1 dimensions, we have also constructed a little group covariant spinor-helicity formalism for $SU(1,1)$ valued-spinors, where 4D $SU(2)$ massive spinors are simply mapped to 3D $U(1)\oplus \overline{U(1)} \in SU(1,1)$ spinors in a natural way. Rather pleasingly, this has allowed us to look at the classical-spin limit of certain observables by extending the spin-deformation of on-shell three-particle amplitudes to three-dimensions. This has allowed us to compute the classical impulse in the gauge theory and gravity cases and may be a useful tool to compute similar observables in other theories.

Somewhat surprisingly, we also showed that the Aharonov--Bohm phase could be derived directly from the tree-level on-shell amplitude in the appropriate limit, and that this precisely double copied to its gravitational equivalent. It would be interesting to see what could be derived in four dimensions, where for example the Dirac string and Misner string ought to give rise to an Aharonov-Bohm phase and do appear in tree-level amplitudes involving dyons \cite{Emond:2020lwi,Huang:2019cja}. It is also plausible that we may be able to study the properties of cosmic strings using on-shell technology.

There are many interesting directions to pursue from here, most pressingly the pure double copy aspects of topologically massive gauge theory. It may also be instructive to examine supersymmetric generalizations of anyon theories \cite{Hlousek:1990jf} and their double copy, as well as to consider the twistorial origin of the 3D classical double copy (both massive and massless) as has recently been examined in four dimensions \cite{White:2020sfn,Chacon:2021wbr,Farnsworth:2021wvs}. We leave these intriguing possibilities to the future. 

\section*{Acknowledgements}
We would like to thank Jeff Murugan for useful discussions and related collaboration. NM would like to thank Mariana Carrillo Gonz\'alez and Donal O'Connell for enlightening discussions. NM is supported by STFC grant ST/P0000630/1. WTE is supported by the Czech Science Foundation GACR, project 20-16531Y. Some of our figures were produced with the help of TikZ-Feynman \cite{Ellis:2016jkw}.
\appendix 
\section{Conventions and Identities}\label{conventions}
We work in Minkowski space with signature of $(-,+,+)$, with the Mandelstam variables are defined as
\begin{equation}
	s = -(p_1+p_2)^2,~~~~~t=-(p_1 + p_3)^2,~~~~~u = -(p_1+p_4)^2 \;.
\end{equation}
In the $SU(1,1)$ representation, the 3D momentum bi-spinor is given by
\begin{equation}\label{pmat}
	p_{\alpha\beta} = p_\mu\tilde{\sigma}^\mu_{\alpha\beta} = \begin{pmatrix}
		-p^1 + ip^2 && p^0 \\ 
		p^0 && -p^1 - ip^2
	\end{pmatrix} \;, 
\end{equation}
where $\mu = 0,1,2$, $\det p_{\alpha\beta} = -(-p_0^2+p_1^2+p_2^2) =-m^2$, and the $\sigma$ and $\epsilon$ matrices are given by
\begin{equation}
	\tilde{\sigma}^0_{\alpha\beta} = -\begin{pmatrix}
		0 && 1 \\ 
		1 && 0
	\end{pmatrix} \;,~~~~~\tilde{\sigma}^1_{\alpha\beta} = -\begin{pmatrix}
		1 && 0 \\ 
		0 && 1
	\end{pmatrix} \;,~~~~~\tilde{\sigma}^2_{\alpha\beta} = \begin{pmatrix}
		i && 0 \\ 
		0 && -i
	\end{pmatrix} \;,~~~~~ \epsilon_{\A\B} = - \epsilon^{\A\B} = \begin{pmatrix}
		0 && -1 \\ 
		1 && 0
	\end{pmatrix} \;.
\end{equation}
These are related to the usual Infeld–Van der Waerden symbols by
\begin{equation}
	\tilde{\sigma}^\mu_{\A\B} = \sigma^\mu_{\A\dA}\chi_{~\B}^{\dA}\;,
\end{equation}
where $\chi_{~\B}^{\dA} = \chi_\mu\epsilon^{\dA\dB}\sigma^\mu_{\B\dB} = \epsilon^{\dA\dB}\sigma^3_{\B\dB}$ and $\chi_\mu = (0,0,0,1)$. Note the convention for raising and lowering spinor indices: as usual the Levi--Civita tensor $\epsilon^{\A\B}$ serves as the metric on spinor space, such that, for example
\begin{equation}
	\tilde{\sigma}^{\mu,\A\B} \ = \ \epsilon^{\A\gamma}\epsilon^{\B\delta}\,\tilde{\sigma}^\mu_{\gamma\delta}\,,\qquad \epsilon_{\A\gamma}\epsilon^{\gamma\B} \ = \ \delta^{~~\B}_{\A}\,,\qquad \epsilon^{\A\B} \ = \ - \epsilon^{\A\gamma}\epsilon^{\beta\delta}\epsilon_{\gamma\delta} \;.
\end{equation}
The gamma matrices are then found by raising the last index, i.e. $(\gamma^\mu)_{\A}^{~~\B} = \epsilon^{\beta\gamma}\tilde{\sigma}^\mu_{\A\gamma}$, such that
\begin{equation}
	(\gamma^0)_{\A}^{~~\B} = \begin{pmatrix}
		-1 && 0 \\ 
		0 && 1
	\end{pmatrix}\;,~~~~~(\gamma^1)_{\A}^{~~\B} = \begin{pmatrix}
		0 && 1 \\ 
		-1 && 0
	\end{pmatrix}\;,~~~~~(\gamma^2)_{\A}^{~~\B} = -\begin{pmatrix}
		0 && i \\ 
		i && 0
	\end{pmatrix}\;.
\end{equation}
These are manifestly traceless $\Tr{\gamma^\mu} = 0$, and satisfy the algebra 
\begin{align}\label{gammaalgebra}
	\gamma^\mu\gamma^\nu &= -\eta^{\mu\nu} - i\epsilon^{\mu\nu\rho}\gamma_\rho\;, \nn\\
	\gamma^\mu\gamma^\nu\gamma^\rho 
	&= \eta^{\mu\rho}\gamma^\nu - \eta^{\rho\nu}\gamma^\mu - \eta^{\mu\nu}\gamma^\rho +i\epsilon^{\mu\nu\rho}\;,
\end{align}
from which one can derive various trace identities, e.g.
\begin{equation}
	\Tr{\gamma^\mu\gamma^\nu} = -2\eta^{\mu\nu} \;,~~~~~\Tr{\gamma^\mu\gamma^\nu\gamma^\rho} = 2i\epsilon^{\mu\nu\rho} \;.
\end{equation}
To avoid any ambiguities in the compact notation, we state here explicitly our conventions for products of spinors:
\begin{flalign}
	\braket{\lambda\bar{\lambda}} \ \equiv \lambda^\A\bar{\lambda}_\A =& \ \epsilon^{\A\B}\lambda_{\B}\bar{\lambda}_\A
	\;,
	\nn\\[-0.6em] \\
	\bra{\lambda}p_1p_2\cdots p_n\ket{\bar{\lambda}} \ =& \ \lambda^\A\,(p_1)_{\A}^{~~\B_1}(p_2)_{\B_1}^{~~\B_2}\cdots(p_n)_{\B_{n-1}}^{~~\B_n}\,\bar{\lambda}_{\B_n} \;, \nn
\end{flalign}
where $(p_i)_{\A}^{~\B}=p_{i\mu}(\gamma^\mu)_{\A}^{~\B}$ .

Using these, we can derive the spinor-helicity identities
\begin{align}\label{fourident}
	\braket{a|p_1p_2|b} &= -\braket{ab}(p_1\cdot p_2) - i\epsilon^{\mu\nu\rho}p_{1\mu}p_{2\nu}\braket{a|\gamma_\rho|b} \;,
	\nn\\[-0.7em] \\
	\braket{a|p_1p_2p_3|b} &= i\braket{ab}\epsilon(p_1,p_2,p_3) + (p_1\cdot p_3)\braket{a|p_2|b} - (p_1\cdot p_2)\braket{a|p_3|b} - (p_2\cdot p_3)\braket{a|p_1|b} \;
	.\nn
\end{align}
For $b = \bar{a}$, this simplifies to become
\begin{align}\label{fourident}
	\braket{a|p_1p_2|\bar{a}} &= \sqrt{-p_a^2}\,(p_1\cdot p_2) + i\epsilon(p_{1},p_{2},p_{a})\;,\nn\\[-0.7em] \\
	\braket{a|p_1p_2p_3|\bar{a}} & \ = \ -i\sqrt{-p_a^2}\,\epsilon(p_1,p_2,p_3) - (p_1\cdot p_3)(p_2\cdot p_a) + (p_1\cdot p_2)(p_3\cdot p_a) + (p_2\cdot p_3)(p_1\cdot p_a)\;.\nn
\end{align}
We decompose the bi-spinor as $p_{\alpha\beta} = \lambda_{\A}\bar{\lambda}_{\B}+\lambda_{\B}\bar{\lambda}_{\A}$, where $\lambda_\A$ and $\bar{\lambda}_\A$ satisfy the following Dirac equations
\begin{equation}
	p_j\ket{j} = -m_j\ket{j}\;,~~~~~p_j\ket{\bar{j}} = m_j\ket{\bar{j}}\;,~~~~~\bra{j}p_j = m_j\bra{j}\;,~~~~~\bra{\bar{j}}p_j = -m_j\bra{\bar{j}}\;.
\end{equation}

These forms allow us to write 
\begin{equation}\label{unsym1}
	\lambda_\A\bar{\lambda}_\B = \frac{1}{2}\left(p_{\A\B} - m\epsilon_{\A\B}\right)\;,~~~~~\bar{\lambda}_\A\lambda_\B = \frac{1}{2}\left(p_{\A\B} + m\epsilon_{\A\B}\right)\;,
\end{equation}
which immediately implies that $ \lambda_{\A}\bar{\lambda}_{\B}+\lambda_{\B}\bar{\lambda}_{\A} = p_{\alpha\beta}$ and that
\begin{align}
	\epsilon^{\alpha\beta}\lambda_{\beta}\bar{\lambda}_{\alpha}= \braket{\lambda\bar{\lambda}} &= -\braket{\bar{\lambda}\lambda} = -m\;.
\end{align}
We can contract in a $\sigma$ matrix to find that $\lambda_\alpha\bar{\lambda}_\beta\sigma^{\mu\alpha\beta} = \frac{1}{2}p_{\alpha\beta}\sigma^{\mu\alpha\beta} - \frac{m}{2}\Tr{\gamma^\mu}$ and therefore that
\begin{equation}
	\braket{i|\gamma^\mu|\bar{i}} = \braket{\bar{i}|\gamma^\mu|i} = -p^\mu\;.
\end{equation}
Using the above relations we can derive the following identities
\begin{equation}
	4\braket{i\bar{j}}\braket{\bar{i}j} = s_{ij} - (m_i-m_j)^2\;,~~~~~~4\braket{\bar{i}\bar{j}}\braket{ij} = s_{ij} - (m_i+m_j)^2\;,
\end{equation}
with $s_{ij} = -(p_i+p_j)^2$.

Throughout this text, we use the Levi-Civita \textit{tensor} in Minkowski space, defined by $\epsilon^{012} = \epsilon_{012} = +1$, with the relationship between upper and lower given by
\begin{equation}
	\epsilon^{\alpha\beta\gamma}\,\eta_{\alpha\kappa}\,\eta_{\beta\lambda}\,\eta_{\gamma\mu} \,=\, \epsilon_{\kappa\lambda\mu} \;,
\end{equation}
and many identities used in the main text can be derived from the relation
\begin{equation}
	\epsilon^{\mu\nu\rho}\epsilon_{\alpha\beta\gamma} = -\begin{vmatrix}
		\delta^\mu_\alpha & \delta^\mu_\beta & \delta^\mu_\gamma \\ 
		\delta^\nu_\alpha & \delta^\nu_\beta & \delta^\nu_\gamma  \\ 
		\delta^\rho_\alpha & \delta^\rho_\beta & \delta^\rho_\gamma
	\end{vmatrix} \;. 
\end{equation}
We note that the relationship between the tensor $\epsilon$ and the symbol $\varepsilon$ is, given our conventions,
\begin{equation}
	\epsilon^{\mu\nu\rho} = \varepsilon^{\mu\nu\rho}\;,~~~~~\epsilon_{\mu\nu\rho} = -\varepsilon_{\mu\nu\rho} \;.
\end{equation}

\section{Lorentz Covariant Massive Particle Spinor Helicity in 2+1 Dimensions}\label{massiveSH3D}
In this appendix, we give a brief introduction to the little--group covariant massive spinor--helicity formalism in $2+1$ dimensions. We do so in direct analogy to the initial comprehensive study by Arkani-Hamed, Huang and Huang  \cite{Arkani-Hamed:2017jhn} in the $3+1$ dimensional case. 

%In the modern era, amplitudes are often expressed using spinor-helicity variables, exchanging vectors and polarizations for spinors and helicity. This has the benefit of making certain symmetries and features manifest, of which can be used to fix many aspects of the amplitudes directly\footnote{For example the three-particle amplitude in four dimensions is entirely fixed by dimensional analysis, little-group scaling and Poincar\'e invariance.}. The variables themselves are constructed by considering local isomorphisms of the Lorentz group, which we can utilise to map tensors with spacetime indices to spinors. 
In $3+1$ dimensions, the vector representation of the Lorentz group is locally isomorphic to $(1/2,1/2)$ or $SL(2)\otimes SL(2)$, such that we can express momenta in terms of bi-spinors: $p_\mu \mapsto p_{\A\dot{\A}} = p_\mu\sigma^\mu_{\A\dot{\A}}$. As usual, the `dotted' indices are to remind us that these objects are chiral, i.e. `dotted' indices have negative chirality (transforming under the $(1/2,0)$ representation), whereas `undotted' have positive chirality (transforming under the $(0,1/2)$ representation). If the momentum is massless, one can write this as the outer product of two Weyl spinors (one from each chiral sector) $\lambda_{\A}$ and $\tilde{\lambda}_{\dot{a}}$ :
\begin{equation}\label{eq:massless 4D bispinor}
	p^2 = -\det(p_{\A\dot{\A}})= 0~~~~~\implies~~~~~ p_{\A\dot{\A}} = \lambda_\A\tilde{\lambda}_{\dot{\A}} = \begin{pmatrix}
		-p^0+p^3 && p^1-ip^2 \\ p^1+ip^2 && -p^0-p^3
	\end{pmatrix} \;.
\end{equation}
One can readily solve the system of equations for the components of $\lambda_\A$ and $\tilde{\lambda}_{\dot{\A}}$, such that they have the particular solutions
\begin{equation}
	\lambda_\A = \frac{1}{\sqrt{-p^0+p^3}}\begin{pmatrix}
		-p^0+p^3 \\ ~~~\,p^1+ip^2
	\end{pmatrix} \;, \qquad \tilde{\lambda}_{\dot{\A}} = \frac{1}{\sqrt{-p^0+p^3}}\begin{pmatrix}
	-p^0+p^3 \\ ~~~\,p^1-ip^2
\end{pmatrix}\;.
\end{equation} 
Note that the corresponding little group for massless particles is $U(1)$. For massive particles the situation is somewhat different: the little group is $SU(2)$ and the determinant of the corresponding momentum bi--spinor $p_{\A\dot{\A}}$ is non--zero. In particular, this means that one cannot express $p_{\A\dot{\A}}$ as a simple outer--product of two Weyl spinors as in the massless case. Nevertheless, at least in the $3+1$  dimensional case, there does exist a prescribed method to describe massive momentum bi--spinors in a little--group covariant manner, thanks to the massive formalism developed in \cite{Arkani-Hamed:2017jhn}.

In $2+1$ dimensions the situation is similar to the four--dimensional case, except now the Lorentz group has multiple isomorphisms:
\begin{equation}
	SO(2,1) \simeq SL(2,\R) \simeq SU(1,1) \simeq SP(2,\R)~.
\end{equation}
Note that (unlike in 4D) in three dimensions the left-- and right--handed chiral representations of the Lorentz group  are \emph{equivalent}, and therefore the distinction between the `dotted' and `undotted' is lost. Moreover, from the perspective of dimensional reduction (to get from a 4D to a 3D representation), the above isomorphisms can be interpreted as different ways of reducing from 4D to 3D, i.e. each isomorphism has a different basis of Pauli matrices. 

Here we concern ourselves with massive particle states, describing their corresponding momenta in terms of spinor-helicity variables. The corresponding little group in this case is $SO(2)\simeq U(1)$. From a dimensional reduction perspective, this is a natural consequence of reducing from 4D to 3D, since this procedure generates a mass for the corresponding 3D momentum bi--spinor $p_{\A\B}^{3D}$ from a massless 4D bi--spinor $p^{4D}_{\A\dot{\A}}$.  With this in mind,  we advocate that $SU(1,1)$ is the natural group to encode the spin of a given particle. The reason being that its little group, defined via 
\begin{equation}\label{eq:little group}
	K_{\A}^{~~\gamma}\,p_\gamma^{~~\delta}\,(K^\dagger)_\delta^{~~\beta} \ = \ p_{\alpha}^{~~\beta} \;,
\end{equation}
encodes both positive and negative helicity modes, since
\begin{equation}
	K = e^{i\theta\,A} = e^{i\theta}\oplus e^{-i\theta}\;,
\end{equation}
where $A^\A_{~\B}$ is the centre of $SU(1,1)$, defined as
\begin{equation}\label{eq:def matrix su11}
	A \ = \ \begin{pmatrix}
		1 && 0 \\ 0 && -1
	\end{pmatrix} \;.
\end{equation}
Hence, $SU(1,1)$ describes spin coherent particle states, as is the case for $SU(2)$ (this is to be expected, since both $SU(1,1)$ and $SU(2)$ are subsets of $SL(2,\mathbb{C})$), with the corresponding spinors transforming under the little group, as 
\begin{equation}
	\lambda_\A \rightarrow e^{-i\theta}\lambda_\A\;,\qquad \bar{\lambda}_\A \rightarrow e^{+i\theta}\bar{\lambda}_\A\;,\qquad p_{\A\B}\rightarrow p_{\A\B} \;,
\end{equation}
such that $p_{\A\B}$ is invariant (as required).

Let us start by constructing momentum $2+1$ dimensional bi--spinors in a particular basis, with the aim of formulating a covariant approach. With this in mind, we consider a basis for $SU(1,1)$ of the form
\begin{equation}
	\tilde{\sigma}^0_{\alpha\beta} = -\begin{pmatrix}
		0 && 1 \\ 
		1 && 0
	\end{pmatrix}\;,~~~~~\tilde{\sigma}^1_{\alpha\beta} = -\begin{pmatrix}
		1 && 0 \\ 
		0 && 1
	\end{pmatrix}\;,~~~~~\tilde{\sigma}^2_{\alpha\beta} = \begin{pmatrix}
		i && 0 \\ 
		0 && -i
	\end{pmatrix}\;,
\end{equation}
such that a 3D momentum bi--spinor is given by
\begin{equation}\label{eq:3d bispinor}
	p_{\A\B} = p_\mu\tilde{\sigma}^\mu_{\A\B} = \begin{pmatrix}
		-p^1 + ip^2 && p^0 \\ 
		p^0 && -p^1 - ip^2
	\end{pmatrix}\;.
\end{equation} 
Since we are taking $p_{\A\B}$ to be generically massive, it follows that 
\begin{equation}
	\text{det}(p_{\A\B}) = p^2 = -m^2 \;. 
\end{equation}
Equation~\eqref{eq:3d bispinor} can then be expressed in terms of an outer--product of two spinors, viz.
\begin{equation}
	p_{\A\B} = \lambda_{\A}\,\bar{\lambda}_{\B} + \lambda_{\B}\,\bar{\lambda}_{\A} \;,
\end{equation}
which admits solutions for $\lambda_{\A}$ and $\bar{\lambda}_{\A}$ of the form
\begin{equation}
	\lambda_{\A} = -\frac{i}{\sqrt{2(p^0+m)}}\begin{pmatrix}
		~p^0+m \\ -p^1-ip^2
	\end{pmatrix}\;,\qquad \bar{\lambda}_{\A} = \frac{i}{\sqrt{2(p^0+m)}}\begin{pmatrix}
	 -p^1+ip^2 \\ ~p^0+m
\end{pmatrix} \;.
\end{equation}
It follows directly from this, that $\lambda_{\A}$ and $\bar{\lambda}_{\A}$ satisfy the inner--product
\begin{equation}
	\braket{\lambda\bar{\lambda}} = \ -m \;,
\end{equation}
and furthermore that
\begin{equation}
	\bra{\lambda}\gamma^\mu\ket{\bar{\lambda}} = \bra{\bar{\lambda}}\gamma^\mu\ket{\lambda} =  \frac{1}{2}p_{\A\B}\tilde{\sigma}^{\mu,\A\B} = -p^\mu \;.
\end{equation}
At this juncture, it is instructive to note (as mentioned earlier) that one can arrive at this construction of massive 3D momentum bi--spinors via dimensional reduction of \emph{massless} 4D bi--spinors. This also serves as a consistency check of the results obtained in the previous appendix. 

Indeed, let us start out in $3+1$ dimensions and consider strictly massless momentum bi--spinors $p^{4D}_{\A\dot{\A}}$ (cf.~\eqref{eq:massless 4D bispinor}). Assuming we wish to keep the momentum Lorentzian, there are three directions we can dimensionally reduce along. Here we shall do so along the $p_3$ direction, which as we shall see, corresponds to  describing the resulting 3D massive momentum vectors in the $SU(1,1)$ representation. We can encode this process in a bi--spinor operator acting on a massless 4D momentum bi--spinor, projecting out its $z$--component. This can be achieved by considering a four--vector of the form $\chi^\mu=(0,0,0,1)$, from which we can construct the bi--spinor
\begin{equation}
	\chi^{\dot{\A}}_{\,\,\A} \ = \ \epsilon^{\dot{\A}\dot{\B}}\,\sigma^3_{\A\dot{\B}} = -\begin{pmatrix}
		0 && 1 \\ 1 && 0
	\end{pmatrix}\;,
\end{equation}
where $\sigma^3$ is the standard third Pauli matrix.

Given this, we find 
\begin{equation}
	p^{4D}_{\A\dot{\A}}\chi^{\dot{\A}}_{\,\,\B} \ = \ p^{4D}_{(\A\dot{\A}}\chi^{\dot{\A}}_{\,\,\B)} + p^3\epsilon_{\A\B} \;,
\end{equation}
from which, we have (using the properties of the determinant)
\begin{equation}
	0= -\text{det}(p^{4D}_{\A\dot{\A}}) \ = \ \text{det}(p^{4D}_{(\A\dot{\A}}\chi^{\dot{\A}}_{\,\,\B)}) + (p^3)^2 \;.
\end{equation}
We note here, that $p^3$ serves as a free parameter in the dimensional reduction, and is chosen so as to match up with the direct calculation in 3D. Motivated by our earlier analysis, if we set $p^3=-m$, with $m$ corresponding to the mass of the 3D momentum, then we can make the natural identification 
\begin{equation}\label{eq: 3d bispinor dim red}
	p^{3D}_{\A\B} = p^{4D}_{(\A\dot{a}}\chi^{\dot{a}}_{\,\,\B)} = \lambda_{\A}\,\bar{\lambda}_{\B} + \lambda_{\B}\,\bar{\lambda}_{\A} \coloneqq p_{\A\B}  \;,
\end{equation}
such that 
\begin{equation}
	\text{det}(p^{3D}_{\A\B}) + m^2 \ = \ (p^{3D})^2 + m^2 \ = \ 0 \quad\longrightarrow\quad \text{det}(p^{3D}_{\A\B}) \ = \ (p^{3D})^2 \ = \ -m^2 \;.
\end{equation}
Note from eq.~\eqref{eq: 3d bispinor dim red} that the determinant can also be expressed as $\text{det}(p^{3D}_{\A\B})=-\braket{\lambda\bar{\lambda}}^2$, which implies that $\braket{\lambda\bar{\lambda}}^2=m^2$. This leaves us with a potential freedom to choose the sign of $\braket{\lambda\bar{\lambda}}$, however, to match with the previous analysis directly in 3D, we are required to choose
\begin{equation}\label{eq:dim red inner prod}
	\braket{\lambda\bar{\lambda}} \ = \ -m \ = \ -\braket{\bar{\lambda}\lambda} \;.
\end{equation}
Moreover, in general, we can express a given bi--spinor $\psi_{\A\B}$ as 
\begin{equation}
	\psi_{\A\B} \ = \ \psi_{(\A\B)} + \frac{1}{2}\,\psi^{\gamma}_{~~\gamma}\,\epsilon_{\A\B} \;,
\end{equation}
where $\psi^{\A}_{~~\A}\coloneqq \epsilon^{\A\B}\psi_{\B\A}$. 
It immediately follows from this, along with eqs.~\eqref{eq: 3d bispinor dim red} and~\eqref{eq:dim red inner prod}, that 
\begin{equation}
	\lambda_{\A}\,\bar{\lambda}_{\B} = \frac{1}{2}(p_{\A\B}-m\epsilon_{\A\B}) \;, \qquad \bar{\lambda}_{\A}\,\lambda_{\B} = \frac{1}{2}(p_{\A\B}+m\epsilon_{\A\B}) \;,
\end{equation}
Observe that since $\chi^{\dot{a}}_{\,\,\A}$ only acts on right--chiral spinors, we can relate the 3D spinor helicity variables to their 4D counterparts as
\begin{equation}\label{eq: 4dtosu11 spinors}
	\lambda^{3D}_{\A} \ = \ \frac{1}{\sqrt{2}}\lambda^{4D}_\A\Big|_{p_3 = -m}\;,~~~~~~~~\bar{\lambda}^{3D}_{\A} \ = \ \frac{1}{\sqrt{2}}\tilde{\lambda}^{4D}_{\dot{a}}\,\chi^{\dot{a}}_{~\A}\Big|_{p_3 = -m}\;.
\end{equation}
Moreover, one can show that $P^\A_{~\B}\coloneqq\frac{1}{\sqrt{-m^2}}p^\A_{~\B}\in SU(1,1)$ (with $p^\A_{~\B}=\epsilon^{\A\gamma}p_{\gamma\B}$), i.e. $P$ satisfies
\begin{equation}
	P^\dagger A\,P \ = \ A \;,\qquad \text{det}(P) \ = \ 1 \;,
\end{equation}
where $A$ is given by eq.~\eqref{eq:def matrix su11}.

Thus, by dimensionally reducing from a massless 4D momentum bi--spinor along the $z$--direction, we have naturally arrived at an $SU(1,1)$ representation of a massive 3D momentum. This is a nice confirmation of the results we found by constructing massive momentum bi-spinors directly in $SU(1,1)$ in $2+1$ dimensions.

This is all well and good, but so far, we have restricted ourself to a particular choice of helicity basis. However, as promised earlier, we can go further in our analysis, and construct little--group covariant spinors, in analogy to the 4D case~\cite{Arkani-Hamed:2017jhn}. We shall adopt capitalised Latin letters $I,J,\ldots$ for the little group (in this case $SO(2)\simeq U(1)$) indices, and keep Greek letters $\A,\B,\ldots$ for the spinor indices. 

We therefore define little--group covariant spinors via the momentum bi--spinor as
\begin{equation}
	p_{\A\B} = \lambda_\alpha^{~~I}\tilde{\lambda}_{\beta I} = \lambda_{\A}\,\bar{\lambda}_{\B} + \lambda_{\B}\,\bar{\lambda}_{\A}\;,
\end{equation}
The spinors can be expanded in a basis $\lbrace\xi^{+I},\xi^{-I}\rbrace$ of the little group as
\begin{flalign}
	\lambda_\alpha^{~~I} \ =& \ \lambda_\alpha\,\xi^{+I} + \bar{\lambda}_\alpha\,\xi^{-I} \;,\nn\\[0.5em] \tilde{\lambda}_\alpha^{~~I} \ =& \ \bar{\lambda}_\alpha\,\xi^{-I} - \lambda_\alpha\,\xi^{+I} \;,
\end{flalign}
where the basis vectors are defined via
\begin{equation}
	\epsilon_{IJ}\,\xi^{-I}\xi^{+J} \ = \ 1\;, \qquad \xi^{+I}\xi^{-J} - \xi^{-I}\xi^{+J} \ = \ \epsilon^{IJ} \;.
\end{equation}
Note that $\lambda_\alpha^{~~I}$ and $\tilde{\lambda}_\alpha^{~~I}$ are related via
\begin{equation}
	\tilde{\lambda}_\alpha^{~~I} \ = \ A^{IJ}\lambda_{\alpha J}\;,
\end{equation}
where $A_J^{~~I}$ is given by eq.~\eqref{eq:def matrix su11}, and $A^{IJ}=A^I_{~L}\epsilon^{LJ}$.

Given these relations, we can readily determine that $\lambda_\alpha^{~~I}$ and $\tilde{\lambda}_\alpha^{~~I}$ satisfy the following inner products:
\begin{equation}
	\braket{\lambda^I\lambda^J} \ = \ m\,\epsilon^{IJ} \;,\qquad\braket{\tilde{\lambda}^I\tilde{\lambda}^J} \ = \ -m\,\epsilon^{IJ} \;,\qquad \braket{\lambda^I\tilde{\lambda}^J} \ = \ m(\tilde{\sigma}^0)^{IJ} \;,
\end{equation}
where we have used that $A^{I}_{~K}\epsilon^{KJ}=(\tilde{\sigma}^0)^{IJ}$. 

Using these, we can imply the following $SU(1,1)$ Dirac equations
\begin{equation}
	p_{\A\B}\lambda^{\A I} \ = \ m\tilde{\lambda}_\B^{~~I} \;,\qquad p_{\A\B}\tilde{\lambda}^{\B I} \ = \ m\lambda_\A^{~~I} \;.
\end{equation}

Equipped with this information, we can readily relate the 4D covariant spinor helicity formalism to the 3D version. Indeed, we can rewrite inner products as
\begin{equation}
	\lambda_a^i\lambda_b^j\epsilon^{ab} = \epsilon^{\A\B}\lambda_\B^i\lambda_\A^j\;,~~~~~\epsilon^{\da\db}\tilde{\lambda}_{\db}^i\tilde{\lambda}_{\da}^j = \epsilon^{\da\db}\chi_{~\db}^{\B}\chi_{~\da}^{\A}\bar{\lambda}_\A\bar{\lambda}_\B = -\epsilon^{\A\B}\bar{\lambda}_\A\bar{\lambda}_\B\;,
\end{equation}
where we have used $\epsilon^{\da\db}\chi_{~\da}^{\A}\chi_{~\db}^{\B} = -\epsilon^{\A\B}$. We find then that, in bra-ket notation
\begin{equation}
	\braket{ij}_{4D} = \braket{ij}_{3D}\;,~~~~~[ij]_{4D} = -\braket{\bar{i}\bar{j}}_{3D}\;.
\end{equation}
In four dimensions, the little group indices $I,J,K...$ are also $SU(2)$ indices, which we can treat in exactly the same manner, converting them to $U(1)\oplus \overline{U(1)} \in SU(1,1)$ indices by simply `bolding' the inner products above to find little group covariant expressions
\begin{equation}
	\braket{\textbf{ij}}_{4D} = \braket{\textbf{ij}}_{3D}\;,~~~~~[\textbf{ij}]_{4D} = -\braket{\textbf{ij}}_{3D}\;.
\end{equation}
For the individual spinors, we find that schematically we can replace
\begin{equation}
	 \lambda_{4D,\alpha}^I \rightarrow \lambda_{3D,\alpha}^I\;,~~~~~\tilde{\lambda}_{4D,\dA}^I \rightarrow \tilde{\lambda}_{3D,\alpha}^I\;.
\end{equation}
Thus, we see that by adopting the $SU(1,1)$ representation of massive momenta in three dimensions, we can relate the 3D little-group covariant formalism to the 4D version in a very straightforward manner. 

\section{$x$-ratio expansion}\label{xratcalc}
In three dimensions, as in the four-dimensional massive spinor helicity formalism, we can introduce the so--called $x$--factors, i.e. the constants of proportionality between massless basis spinors when the mass of each external particle are equal. Working in the $SU(1,1)$ representation, these are given by
\begin{equation}\label{eq:x variables}
	x_i \ = \ \frac{\bra{q}u_i\ket{\eta}}{\braket{\bar{q}\eta}} \;, \qquad \frac{1}{x_i} \ = \ \frac{\bra{\bar{q}}u_i\ket{\bar{\eta}}}{\braket{q\bar{\eta}}} \;, 
\end{equation} 
where $q^\mu$ is the transfer momentum (at the vertex), $\eta^\mu$ is an arbitrary reference vector (that we can freely choose), and $u_i^\mu=\frac{p_i^\mu}{m_i}$ is the three--velocity of the external particle $i$. 

In this paper, we concern ourselves with the case of a massive internal particle. We will therefore use the gauge-fixed massive $x$-factor (derived in \cite{Moynihan:2020ejh}), given by
\begin{equation}\label{eq:massive x factor}
	x_i = \frac{\bra{q}u_i\ket{q}}{\braket{\bar{q}q}},~~~~~\frac{1}{x_i} = \frac{\bra{\bar{q}}u_i\ket{\bar{q}}}{\braket{q\bar{q} }} \;.
\end{equation}
We note here, that unlike in the massless case, $x_i$ and $\frac{1}{x_i}$ are not strictly inverses of one another. The is not technically a problem in itself, as they need not be inverses in general. However, in the regime that we are interested in here, i.e. long--range interactions, they are essentially inverse to one another, hence why we have kept the standard suggestive notation. Indeed, one can show that $\frac{1}{x_i}x_i = 1 + \frac{q^2}{4m_i^2}$, and for all cases we consider here $q^2\ll m_i^2$, such that \smash{$\frac{1}{x_i}x_i \simeq 1$}.
 
With this in mind, using eq.~\eqref{eq:massive x factor} along with eq.~\eqref{unsym1} and the relations for products of spinors with $\gamma$--matrices given in eq.~\eqref{fourident}, we find
\begin{align}
	\frac{x_i}{x_j}&=\frac{\braket{q|u_i|q}\braket{\bar{q}|u_j|\bar{q}}}{q^2} \nn\\
	&= \frac{\braket{q|u_iqu_j|\bar{q}} + m\braket{q|u_iu_j|\bar{q}}}{2q^2} \nn\\
	&= \frac{2(u_i\cdot q)(u_j\cdot q) - 2q^2(u_i\cdot u_j) + 2m\epsilon(u_i,u_j,q)}{2q^2} \nn\\
	&= -\frac{q^2(u_i\cdot u_j) -(u_i\cdot q)(u_j\cdot q) - im\epsilon(u_iu_jq)}{q^2}\;,
\end{align}
It is useful from the perspective of the double copy to express this in terms of the rapidity $w$. Since we are interested in long range interactions, we can discard any terms that go like $\cl{O}(q^2)$, especially since these will enter as $\frac{m^2}{m_1m_2}$. In this case, we can express the massive particle $x$-ratio as
\begin{align}
	\frac{x_1}{x_2} &= -u_1\cdot u_2 +\frac{(u_1\cdot q)(u_2\cdot q)}{q^2}\pm \frac{im}{q^2}\sqrt{-\Gamma(u_1,u_2,q)} \nn\\
	&= -u_1\cdot u_2 - \frac{m^2}{4m_1m_2} \pm \sqrt{\sinh^2w + \frac{m^2}{4m_1^2m_2^2}\left(m_1^2 + m_2^2 - 2m_1m_2\cosh w\right)} \nn\\
	&= \cosh w \pm \sinh w + \cl{O}\left(\frac{m^2}{m_1m_2}\right)\;,
\end{align}
where we have used $u_1\cdot u_2 = -\cosh w$ and $\sinh^2w = (u_1\cdot u_2)^2 - 1$. In the four dimensional case, the two $\pm$ solutions to the $x$-ratio was related to the Dirac string. In this case, it is related to the choice of sign for the Chern-Simons level number.

It will be useful to express $\sinh w$ in another way when considering e.g. the classical impulse. We can expand an arbitrary vector $k^\mu$ into a basis given by 
\begin{equation}
	k^\mu = \alpha\epsilon^\mu(u_1,u_2) + \beta \epsilon^\mu(u_1,q) + \gamma \epsilon^\mu(u_2,q) \;.
\end{equation}
Solving for the coefficients, we find  
\begin{equation}
	k^\mu = \frac{k\cdot q}{\epsilon(q,u_1,u_2)}\epsilon^\mu(u_1,u_2) - \frac{k\cdot u_2}{\epsilon(q,u_1,u_2)} \epsilon^\mu(u_1,q) + \frac{k\cdot u_1}{\epsilon(q,u_1,u_2)} \epsilon^\mu(u_2,q) \;.
\end{equation}
Importantly for our purposes, this allows us to express $q^\mu$ as
\begin{align}\label{qsinh}
	q^\mu &= -\frac{m^2}{\epsilon(q,u_1,u_2)}\epsilon^\mu(u_1,u_2) + \frac{m^2}{2m_1m_2}\frac{1}{\epsilon(q,u_1,u_2)}\Big[m_1\epsilon^\mu(u_1,q) - m_2\epsilon^\mu(u_2,q)\Big] \nn\\
	&= -\frac{m^2}{\epsilon(q,u_1,u_2)}\epsilon^\mu(u_1,u_2) + \cl{O}\left(\frac{m^2}{m_1m_2}\right) \;.
\end{align}
Crucially, this means we can write
\begin{equation}
	q^\mu \sinh w = -m\epsilon^\mu(u_1,u_2) + \cl{O}\left(\frac{m^2}{m_1m_2}\right)\;.
\end{equation}

\section{Spin Deformations of Three-Particle Amplitudes}\label{app:spindef}
In this appendix we will derive the spin-deformation of the three-particle amplitude as used in an earlier section.
While the spin is characterised by a pseudoscalar in $2+1$ dimensions, there is a vector that we can construct, the magnetic moment vector (for spin $1/2$), given by \cite{Plyushchay:1991py}
\begin{equation}\label{magmoment}
	S^\mu = \frac{\epsilon^{\mu\nu\rho}q_\rho p_{2\nu}}{2m_2} \;.
\end{equation} 

In Ref. \cite{Bautista:2019tdr}, Bautista and Guevara showed that a classical spin-multipole expansion can be obtained by considering a Lorentz boost that relates the incoming and outgoing particle momenta. In order to do this in three dimensions, we use little-group covariant $SU(1,1)$ spinor helicity variables. We consider the outgoing spin-$s$ particle as being boosted by an amount $q^\mu$, which is given by \cite{Jackiw:1990ka}
\begin{equation}
	\ket{\textbf{2}'} = (\mathbb{I} + \omega_{\mu}J^\mu)\ket{\textbf{2}}\;,
\end{equation}
where $\omega_\mu$ is the boost parameter and dual to $\omega_{\mu\nu}$.

Boosting a particle initially at rest by a momentum $q^\mu$ with $q^2 = -m^2$ is given by
\begin{equation}
	\omega^{\mu\nu} = \eta^{\mu\nu} + \frac{(q^\mu + mu^\mu)(q^\nu + mu^\nu)}{m(q\cdot u + m)} - 2\frac{q^\mu u^\nu}{m}\;,
\end{equation}
where $u^\mu = (1,0,0)$.

Writing momentum $p_2^\mu = (m_2,0,0) = m_2 u^\mu$, we can express this as
\begin{align}
	\omega^{\mu\nu} &= \eta^{\mu\nu} + \frac{(m_2q^\mu + mp^\mu_2)(m_2q^\nu + mp_2^\nu)}{m_2m(q\cdot p_2 + m_2m)} - 2\frac{q^\mu p^\nu_2}{mm_2}\;.
\end{align}

We can then write down $\omega_\mu$, expressed as the dual of $\omega_{\mu\nu}$ as
\begin{equation}
	\omega_\mu = \frac12\epsilon_{\mu\nu\rho}\omega^{\nu\rho} = -\frac{\epsilon_{\mu\nu\rho}q^\nu p_2^\rho}{mm_2}\;,
\end{equation}
meaning we can express the boost of a single spinor as
\begin{equation}
	\ket{\textbf{2}'} = \left(1 + \frac{1}{2mm_2}\epsilon_{\mu\nu\rho}p_2^\mu q^\nu\gamma^\rho\right)\ket{\textbf{2}}\;.
\end{equation}
We then recognise the spinor contraction as containing the magnetic moment vector, i.e.
\begin{equation}\label{spin-vector}
	\frac{1}{m_2}\braket{\textbf{22}'} = \mathbb{I}+\frac{\braket{\textbf{2}|S\cdot J|\textbf{2}}}{m}\;, 
\end{equation}
where we now define
\begin{equation}
	\frac{1}{m}(S\cdot J)_{\alpha_1\cdots\alpha_{2s}}^{\qquad\beta_1\cdots\beta_{2s}} \ = \ -\frac{1}{2mm_2}\epsilon_{\mu\nu\rho}q^\mu p_2^\nu \big(J^{\rho}\big)_{\alpha_1\cdots\alpha_{2s}}^{\qquad\beta_1\cdots\beta_{2s}} \ \cong \ \frac{2s}{m}\,(S_{1/2}\cdot J)_{\alpha_1}^{\;\;\beta_1}\,\bar{\mathds{I}}_{1}\;.
\end{equation}
with
\begin{equation}\label{pseudovector}
	\frac{1}{m}(S_{1/2}\cdot J)^\alpha_\beta = \left(\frac{1}{mm_2}\epsilon_{\mu\nu\rho}q^\mu p_2^\nu J^\rho\right)^\alpha_\beta\;.
\end{equation}

Since it arises naturally from the Lorentz boost, we will use $S^\mu$ to characterise the spin from here on. 

Using the group algebra, and expressing the internal momentum dependence in terms of its associated wavenumber $\bar{q} = \hbar q$, we can express the dot product as
\begin{equation}
	\frac{1}{m}(S_{1/2}\cdot J)^\A_\B = \frac{1}{2mm_2}\epsilon_{\mu\nu\rho}q^\mu [p_2^\nu,J^{\rho}]^\A_\B = \hbar\frac{a_2\cdot \bar{q}}{m}\delta^\A_\B = \frac{\hbar}{2}\frac{m}{m_2}\delta^\A_\B = \frac{s\hbar m}{m_2}\delta^\A_\B\;. 
\end{equation}
Thus, we can write
\begin{equation}\label{spin-vector2}
	\frac{\braket{\textbf{22}'}^{2s}}{m_2^{2s}} = \left(\mathbb{I}+\frac{\braket{\textbf{2}|S_{1/2}\cdot J|\textbf{2}}}{m}\right)^{2s} = \left(\mathbb{I}+\frac{\braket{\textbf{2}|S\cdot J|\textbf{2}}}{2sm}\right)^{2s} = \left(\mathbb{I}+\frac{s\hbar m}{2sm_2}\right)^{2s}\;, 
\end{equation}
which, in the infinite spin limit keeping $s\hbar$ fixed, becomes
\begin{equation}\label{spin-vector3}
	\lim_{2s\rightarrow \infty}\frac{\braket{\textbf{22}'}^{2s}}{m_2^{2s}} = e^{\frac{m}{m_2}\sigma}\;,
\end{equation}
where we have identified the classical spin as $$\sigma = \lim_{s\rightarrow\infty,\hbar\rightarrow 0} s\hbar\;.$$
\section{Fourier Transforms}\label{sec:FT}
In this appendix, we collect some useful results for 2D Fourier transforms that we make use of in the main body. We will choose the convention that the Fourier transform and its inverse are given by
\begin{equation}
	f(x) = \int \frac{\sd^nq}{(2\pi)^n} \tilde{f}(q)e^{-iq\cdot x} = \int \hat{\sd}^nq \tilde{f}(q)e^{-iq\cdot x},~~~~~\tilde{f}(q) = \int \sd^nx f(x)e^{iq\cdot x}\;.
\end{equation}

We first note, that if a function $f(\mathbf{r})$ and its Fourier transform $\tilde{f}(\mathbf{q})$ are purely radial, i.e., $f(\mathbf{r})=f(|\mathbf{r}|)$ and $\tilde{f}(\mathbf{q})=\tilde{f}(|\mathbf{q}|)$, then 
\begin{equation}\label{radial FT}
	f(\mathbf{r}) = \int\hat{\sd}^2\mathbf{q}\,e^{i\mathbf{q}\cdot \mathbf{r}}\,\tilde{f}(\mathbf{q}) = \frac{1}{(2\pi)^2}\int_{0}^{\infty}\sd q \,q\,\tilde{f}(q)\int_{0}^{2\pi}\sd\phi\,e^{iqr\cos(\phi)} = \int_{0}^{\infty}\frac{\sd q}{2\pi} \,q\,\tilde{f}(q)\,J_0(qr)\;,
\end{equation}
where we have used that 
\begin{equation}
	\int_{0}^{2\pi}\sd\phi\,e^{iqr\cos(\phi)} = 2\pi J_0(qr) \;,
\end{equation}
with $J_n(z)$ a Bessel function of the first kind. Furthermore, we have adopted the compact notation $\int\hat{\sd}^2\mathbf{q}\coloneqq \int_{-\infty}^{\infty}\sd^2\mathbf{q}/(2\pi)^2$, and set $|\mathbf{r}|=r$ and $|\mathbf{q}|=q$.

Let us now consider a particular example function $\tilde{f}(q)=\frac{1}{q^2+m^2}$ (where $m$ is a mass parameter) that will be used extensively in the main body. We will consider the two cases $m>0$ and $m=0$. Making use of eq.\eqref{radial FT}, the two cases can be summarised as follows:
%\footnote{Strictly speaking, this result only holds for $r>0$, as can be seen by the fact that $K_0(\mu r)\rightarrow\infty$ when $r\rightarrow 0$ (a priori assuming that $\mu>0$).}
\begin{equation}\label{mu non zero}
	\int\hat{\sd}^2\mathbf{q}\,e^{i\mathbf{q}\cdot \mathbf{r}}\,\frac{1}{q^2+m^2} \ = \ \begin{cases}
		\frac{1}{2\pi}K_0(mr)&\quad m\neq 0\;, \\[0.5em] -\frac{1}{2\pi}\text{log}(r)\,&\quad m = 0 \;,
	\end{cases}
\end{equation}
where $K_n(z)$ is a modified Bessel function of the second kind.

\section{Scalar Amplitudes in $\cl{N}=0$ Supergravity in $D$-Dimensions}\label{app:sugra}
In this appendix, we derive the scalar $2\rightarrow 2$ scattering amplitudes in so-called $\cl{N}=0$ supergravity in general dimension $D$. $\cl{N}=0$ supergravity consists of the graviton, the dilaton and the Kalb-Ramond field (all fields massless) and, in the string frame, is governed by the action
\begin{equation}
	S_{string} = \frac{2}{\kappa_D^2}\int \sd^Dx\sqrt{-\tilde{g}}e^{-2\varphi}\left[R + 4\tilde{g}^{\mu\nu}\pd_\mu\varphi\pd_\nu\varphi - \frac{1}{12}H_{\mu\nu\rho}H^{\mu\nu\rho}\right] \;,
\end{equation}
where $\varphi$ is the dilaton and $H_{\mu\nu\rho}$ the field strength for the antisymmetric Kalb-Ramond field $B_{\mu\nu}$. By performing a conformal transformation of the metric, $\tilde{g}_{\mu\nu}\rightarrow \Omega^2 g_{\mu\nu}$, we can rewrite this in the Einstein frame, including the de Donder gauge fixing term, as
\begin{equation}\label{N0action}
	S = \int \sd^Dx\sqrt{-g}\left[\frac{2}{\kappa_D^2}R -\frac{1}{2(D-2)}g^{\mu\nu}\pd_\mu\varphi\pd_\nu\varphi - \frac{1}{6}e^{-\sqrt{\frac{8}{D-2}\varphi}}H_{\mu\nu\rho}H^{\mu\nu\rho} + \pd_\mu \bar{h}^{\mu\nu}\pd^\rho h_{\rho\nu}\right] \;.
\end{equation}
We recognise each field as being either the symmetric, anti-symmetric or trace of some two-tensor field $H_{\mu\nu}$, i.e.
\begin{equation}
	h_{\mu\nu} = H_{(\mu\nu)}\;,~~~~~B_{\mu\nu} = H_{[\mu\nu]}\;,~~~~~\varphi = H^\mu_{~\mu} \;.
\end{equation}
The polarization tensors for each of these fields are determined by 
\begin{align}
	\epsilon_{\mu\nu}^{(h)} &= \epsilon_{\nu\mu}^{(h)}\;,&&\epsilon_{\mu\nu}^{(h)}\eta^{\mu\nu} = q^{\mu}\epsilon_{\mu\nu}^{(h)} = 0\;, \nn\\
	\epsilon_{\mu\nu}^{(B)} &= -\epsilon_{\nu\mu}^{(B)}\;,&&\epsilon_{\mu\nu}^{(B)}\eta^{\mu\nu} = q^{\mu}\epsilon_{\mu\nu}^{(B)} = 0\;,\\
	\epsilon_{\mu\nu}^{(\phi)} &= \frac{1}{\sqrt{D-2}}(\eta_{\mu\nu} -q_\mu \eta_\nu-q_\nu \eta_\mu)\;,&&q^{\mu}\epsilon_{\mu\nu}^{(\phi)} = 0 \;, \nn
\end{align}
where $\eta_\mu$ is defined via $\eta\cdot q = 1$. At tree-level, coupling to a symmetric, conserved stress-energy tensor, then neither the $B$-field nor the $q_\mu \eta_\nu$ contribute and will be ignored from now on. In the de Donder gauge, the graviton propagator is given by
\begin{equation}\label{gravprop}
	\cl{D}(q)_{\mu\nu\rho\sigma} = \frac{i}{2q^2}\left(\eta_{\mu\rho}\eta_{\nu\sigma}+\eta_{\mu\sigma}\eta_{\nu\rho} - \frac{2}{D-2}\eta_{\mu\nu}\eta_{\rho\sigma}\right)\;,
\end{equation}
while the dilaton propagator by
\begin{equation}\label{gravprop}
	D(q) = \frac{i}{q^2}\frac{\alpha}{D-2}\;,
\end{equation}
where for future convenience we introduce the parameter $\alpha$ to keep track of dilaton contribution, with $\alpha = 1$ corresponding to the action \eqref{N0action}.

The scalar-scalar-graviton vertex is given by
\begin{equation}\label{scalarvertex}
	V^{\mu\nu} = -i\frac{\kappa}{2}\left(p_1^\mu p_2^\nu+p_1^\nu p_2^\mu - \frac{1}{2}\eta^{\mu\nu} q^2\right)\;,
\end{equation}
which we can use to construct the $2\rightarrow 2$ scalar amplitude as mediated by either the graviton or dilaton. 
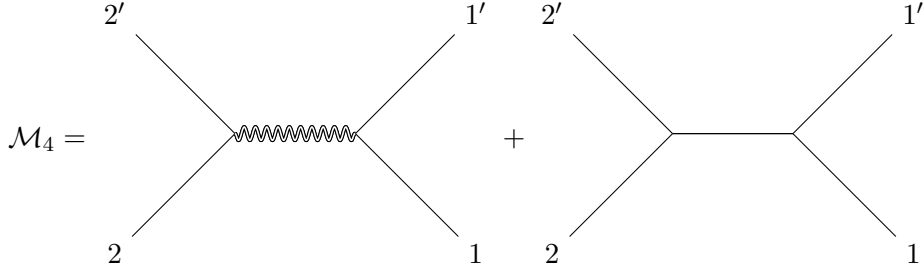
\begin{figure}[H]
	\centering
	\begin{equation}
	\begin{gathered}
		\cl{M}_4
	\end{gathered}
	=
	\begin{gathered}
	\begin{tikzpicture}[scale=0.8]
		\begin{feynman}  
			\vertex (a) at (-4,2) {$2'$};
			\vertex (b) at (-4,-2) {$2$};
			\vertex (c) at (2,-2) {$1$};
			\vertex (d) at (2,2) {$1'$};
			\vertex (r) at (0,0);
			\vertex (l) at (-2,0) ;
			\diagram* {
				(a) -- [plain] (l) -- [graviton] (r) -- [plain] (d),
				(b) -- [plain] (l) -- [graviton] (r) -- [plain] (c),
			};
		\end{feynman}
	\end{tikzpicture}
\end{gathered}
+
\begin{gathered}
	\begin{tikzpicture}[scale=0.8]
		\begin{feynman}  
			\vertex (a) at (-4,2) {$2'$};
			\vertex (b) at (-4,-2) {$2$};
			\vertex (c) at (2,-2) {$1$};
			\vertex (d) at (2,2) {$1'$};
			\vertex (r) at (0,0);
			\vertex (l) at (-2,0) ;
			\diagram* {
				(a) -- [plain] (l) -- [plain] (r) -- [plain] (d),
				(b) -- [plain] (l) -- [plain] (r) -- [plain] (c),
			};
		\end{feynman}
	\end{tikzpicture}
\end{gathered} \nn
\end{equation}
	\caption{Graviton \& dilaton mediated interactions}
	\label{dilatondiags}
\end{figure}
We find that the two diagrams contribute (as in Fig. \ref{dilatondiags}) in the $t\rightarrow 0$ limit
\begin{align}
	\cl{M}_4[1,2\rightarrow 1',2'] &= V(p_1,p_1')^{\mu\nu}\cl{D}_{\mu\nu\rho\sigma}(q)V(p_2,p_2')^{\rho\sigma} + V(p_1,p_1')^{\mu}_{~\mu} D(q)V(p_2,p_2')^{\rho}_{~\rho} \nn\\
	&=  \frac{i\kappa^2 m_1^2m_2^2}{(D-2)q^2}\left((D-2)\sinh^2w + D - 3 + \alpha\right) + \text{contact terms} \;.
\end{align}
We can now compare various scenarios in different dimensions. In $D=4$, the pure graviton contribution gives
\begin{align}
	\cl{M}_4[1,2\rightarrow 1',2']\bigg|_{\alpha=0,~D\rightarrow 4} =  \frac{i\kappa^2 m_1^2m_2^2}{2q^2}\left(2\sinh^2w + 1\right) = \frac{i\kappa^2 m_1^2m_2^2}{2q^2}\cosh2w \;,
\end{align}
which we recognise as the double copy of the charged scalar amplitude when mediated by a photon.

In $D=3$ with $\alpha = 0$, we find a solution which vanishes in the NR limit | the conical singularity solution of pure Einstein gravity
\begin{align}
	\cl{M}_4[1,2\rightarrow 1',2']\bigg|_{\alpha=0,~D\rightarrow 3} =  \frac{i\kappa^2 m_1^2m_2^2}{q^2}\sinh^2w \;.
\end{align}

Finally, we can include both the dilaton and the graviton to find a double copy that holds in any dimension since including the dilaton contribution entirely removes the $D$ dependence of the amplitude
\begin{align}\label{gulamp}
	\cl{M}_4[1,2\rightarrow 1',2']\bigg|_{\alpha=1} =  \frac{i\kappa^2 m_1^2m_2^2}{q^2}\left(\sinh^2w + 1\right) = \frac{i\kappa^2 m_1^2m_2^2}{q^2}\cosh^2w \;.
\end{align}
In general $D$, the origin of this term is a combination of the graviton and the dilaton. This continues to be true in $D=3$, however in this special case it combines the purely topological graviton (the conical singularity solution in Einstein gravity) and the dilaton.

The classical double copy in three dimensions has been studied in detail by Carrillo-Gonz\'alez et al \cite{CarrilloGonzalez:2019gof} (see also \cite{Gumus:2020hbb,Alkac:2021seh}),  where they find that the double copy of a classical point charge gives rise to a gravitational solution with metric
\begin{align}
	\sd s^2 &= -(1+2GM\log r)\sd t^2 + (1+2GM\log r)^{-1}\sd r^2 + r^2\sd\phi^2 \nn\\
		&= -(1+2GM\log r)\sd t^2 + (1-2GM\log r)\sd r^2 + r^2\sd\phi^2 + \cl{O}(G^2) \;.
\end{align}
Taking $M = m_2$, this corresponds to a perturbative metric of the form
\begin{equation}
	h_{\mu\nu}(r) = \text{diag}(-4Gm_2\log{r},-4Gm_2\log{r},0) \;,
\end{equation}
or, in momentum space,
\begin{equation}
	h_{\mu\nu}(q) = -\frac{\kappa^2m_2}{q^2}\text{diag}(1,1,0) \;.
\end{equation}
Plugging this into the impulse, we find
\begin{align}
	\Delta p_1^\mu &= m_1\int \sd^2q\,\delta(u_1\cdot q)\delta(u_2\cdot q)\,e^{iq\cdot b}q^{[\mu}h_{\nu]\rho}u_1^\nu u_1^\rho \nn\\
	&= m_1\int \sd^2q\,\delta(u_1\cdot q)\delta(u_2\cdot q)\,e^{iq\cdot b}(q^{\mu}h_{\nu\rho}u_1^\nu u_1^\rho + q\cdot u_1 h^{\mu\nu}u_{1\nu}) \nn\\
	&= -\kappa^2m_1m_2\cosh^2w\int \sd^2q\,\delta(u_1\cdot q)\delta(u_2\cdot q)\,e^{iq\cdot b}\frac{q^{\mu}}{q^2} \;,
\end{align}
where we have taken the incoming particle to be travelling along the $y$ direction, e.g. $u_1 = (\cosh w, 0, \sinh w)$. This confirms our earlier result, since it exactly matches the impulse derived from the amplitude in eq. \eqref{gulamp} which, as we saw, corresponds to a combination of the conical singularity solution of Einstein gravity plus a dilaton.
\section{Ghost Contributions in 3D Massive Gravity}\label{app:dilaton}
We saw that scattering amplitudes involving the exchange of a topologically massive graviton had a surprising non-zero \textit{massless} limit, even though the massless graviton has no degrees of freedom in 2+1 dimensions. Furthermore, we found that the double copy of topologically massive gauge theory coupled ot matter was not \textit{quite} topologically massive gravity coupled to matter, but has an extra massless mode in the spectrum. This can be traced to the fact that the spectrum of TMG contains a massless ghost, which can in fact contribute to the scattering amplitude, at least when coupled to matter particles with a non-zero mass \cite{Dengiz:2013hka}.
To see this, we will consider a generic tree-amplitude built from the Feynman rules. 

We will consider a propagator given by

\begin{equation}
\begin{aligned}
	D_{\mu \nu, \alpha \beta}=& \frac{-i\xi / 2}{q^{2}}\left(\eta_{\mu \alpha} \eta_{v \beta}+\eta_{\nu \alpha} \eta_{\mu \beta}-2 \eta_{\mu \nu} \eta_{\alpha \beta}\right) \\
	&+\frac{i / 2}{q^{2}+m^{2}}\left(\eta_{\mu \alpha} \eta_{\nu \beta}+\eta_{\nu \alpha} \eta_{\mu \beta}-\eta_{\mu \nu} \eta_{\alpha \beta}\right) \\
	&+\frac{m / 4}{q^{2}+m^{2}} \frac{q^{\gamma}}{q^{2}}\left(\varepsilon_{\mu \alpha \gamma} \eta_{\nu \beta}+\varepsilon_{\nu \alpha \gamma} \eta_{\mu \beta}+\varepsilon_{\mu \beta \gamma} \eta_{\nu \alpha}+\varepsilon_{\nu \beta \gamma} \eta_{\mu \alpha}\right) \;,
\end{aligned}
\end{equation}
where $\xi$ parametrises the contribution of the massless ghost mode, with $\xi = 1$ recovering TMG and $\xi = 1/2$ giving the double copied version.

A generic tree-level amplitude is given by
\begin{align}
	i\cl{A} = -\frac{\xi}{q^2}\left(T^{\mu\nu}\tilde{T}_{\mu\nu} - T\tilde{T}\right) + \frac{1}{q^2+m^2}\left(T^{\mu\nu}\tilde{T}_{\mu\nu} - \frac12T\tilde{T}\right) + \frac{im}{q^2(q^2+m^2)}\left(T^{\mu}_{~~\beta}\varepsilon_{\mu\alpha}(q)\tilde{T}^{\alpha\beta}\right) \;.
\end{align}

There are two possible sources of poles in this amplitude, one massless and one massive, suggesting that there are two propagating particles. Examining this amplitude further, we discover an important fact: the amplitude does not vanish in the massless limit. This can be seen by considering the propagator on its own, since we find that the massless limit gives
\begin{equation}
	D_{\mu \nu, \alpha \beta}\bigg|_{m\rightarrow 0} = \frac{i/2}{q^2}\left[(\xi-1)\left(\eta_{\mu \alpha} \eta_{\nu \beta}+\eta_{\nu \alpha} \eta_{\mu \beta}\right) + (2\xi - 1)\eta_{\mu \nu}\eta_{\alpha \beta}\right] \;.
\end{equation}
For TMG with $\xi = 1$, then, we should expect a generic amplitude to have a non-zero massless limit provided it is sourced by a conserved stress-energy tensor with a non-zero trace, e.g.
\begin{equation}
	T^{\mu\nu}D_{\mu \nu, \alpha \beta}\tilde{T}^{\alpha\beta}\bigg|_{m\rightarrow 0, \xi = 1} = \frac{i}{2}\frac{T\tilde{T}}{q^2} \;.
\end{equation}
For the double copy with $\xi = 1/2$, we find that amplitudes in the massless limit can be of the form
\begin{equation}
	T^{\mu\nu}D_{\mu \nu, \alpha \beta}\tilde{T}^{\alpha\beta}\bigg|_{m\rightarrow 0, \xi = 1/2} = \frac{i}{2}T^{\mu\nu}\tilde{T}_{\mu\nu} \;.
\end{equation}

If we consider the stress energy tensor to be the scalar vertex considered in the previous section, then we find that first term is given by
\begin{equation}
	i\cl{A}_\xi =  \frac{\kappa^2\xi}{2}\frac{m_1m_2(q^2\cosh w + 2m_1m_2\sinh^2w)}{q^2} = \kappa^2\xi\frac{m_1^2m_2^2\sinh^2w}{q^2} + \text{contact} \;,
\end{equation}
which gives the amplitude associated with the conical singularity of pure Einstein gravity. 
The second term is given by
\begin{equation}
	i\cl{A}_{m^2\neq 0} = -\frac{\kappa^2}{32}\frac{4q^2(m_1^2+m_2^2) + 16m_1^2m_2^2\cosh2w + 16q^2m_1m_2\cosh w + q^4}{q^2+m^2} \;.
\end{equation}
If we now consider the massless limit of the sum of these terms at lowest order in $q^2$, we find
\begin{equation}
	i\cl{A}_{m\rightarrow 0}\bigg|_{q^2\rightarrow 0} = \frac{\kappa^2}{2}\frac{m_1^2m_2^2\left[(\xi-1)\cosh 2w - \xi\right]}{q^2} + \text{contact} \;.
\end{equation}
For $\xi = 1$, the value in TMG, we find that the amplitude is that of the dilaton exchange, e.g.
\begin{equation}
	i\cl{A}_{m\rightarrow 0}\bigg|_{q^2\rightarrow 0,\xi = 1} = -\frac{\kappa^2}{2}\frac{m_1^2m_2^2}{q^2} \;,
\end{equation}
whereas for $\xi = 1/2$, the value we get from the double copy, we find that the solution mixes both the massless topological graviton and the dilaton, giving 
\begin{equation}
	i\cl{A}_{m\rightarrow 0}\bigg|_{q^2\rightarrow 0,\xi = 1/2} = -\frac{\kappa^2}{2}\frac{m_1^2m_2^2}{2q^2}\cosh^2w \;,
\end{equation}
which is precisely what we found using the on-shell construction. The origin of these modes was also discussed in \cite{Moynihan:2020ejh}, where the different values of $\xi$ can be understood as the coefficients of the residues of the two poles, either $q^2 = 0$ or $q^2 = -m^2$.

\bibliographystyle{JHEP}
\bibliography{topgrav} 
\end{document}